\documentclass[fleqn,usenatbib,useAMS]{mnras}

\usepackage{newtxtext,newtxmath}

\usepackage[T1]{fontenc}

\DeclareRobustCommand{\VAN}[3]{#2}
\let\VANthebibliography\thebibliography
\def\thebibliography{\DeclareRobustCommand{\VAN}[3]{##3}\VANthebibliography}

\usepackage{graphicx}	
\usepackage{amsmath}	
\usepackage{orcidlink}

\newcommand{\msun}{{\rm M}_{\sun}}
\newcommand{\scl}{Sch{\"o}nberg--Chandrasekhar }

\title[Eccentricities of millisecond pulsars]{Eccentricities of millisecond pulsars with intermediate-mass progenitors}

\author[H. Bareli and S. Ginzburg]{
Hagai Bareli\thanks{\raggedright E-mail: \href{mailto:hagai.bareli@mail.huji.ac.il}{hagai.bareli@mail.huji.ac.il} (HB);
\href{mailto:sivan.ginzburg@mail.huji.ac.il}
{sivan.ginzburg@mail.huji.ac.il} (SG)}\label{emails} and
Sivan Ginzburg$^{\orcidlink{0000-0002-3751-4553}}$\hyperref[emails]{\footnotemark[1]}
\\
Racah Institute of Physics, The Hebrew University, Jerusalem 9190401, Israel
}

\date{Accepted XXX. Received YYY; in original form ZZZ}

\pubyear{\the\year{}}

\begin{document}
\label{firstpage}
\pagerange{\pageref{firstpage}--\pageref{lastpage}}
\maketitle

\begin{abstract}
One channel to form millisecond pulsars with CO white dwarf companions is through the stable Roche-lobe overflow of intermediate-mass ($3\,\msun\lesssim M\lesssim 5\,\msun$) stars at the end of the main sequence (Case A) or the beginning of the hydrogen shell burning phase (Case B). We reproduce previous numerical calculations of this channel and supplement them with a simple analytical model that relates the final orbital period $P(M,m_{\rm wd})$ to the white dwarf's mass and to its progenitor's initial mass $M$. We also theoretically calculate for the first time the eccentricity $e$ in this process, which is set by the fluctuating gravitational quadrupole moment of the progenitor's convective envelope during Roche-lobe detachment. Intermediate-mass progenitors detach when their non-degenerate cores ignite helium, in contrast to low-mass ($M\lesssim 2\,\msun$) stars with degenerate cores that detach when their envelopes become too light to support a burning shell. Despite the order of magnitude higher envelope mass at detachment $m_{\rm e}$ in our case, the eccentricity is barely affected because $e\propto m_{\rm e}^{1/6}$, explaining why intermediate-mass ($m_{\rm wd}\lesssim 0.6\,\msun)$ CO white dwarfs have similar eccentricities to lower mass helium white dwarfs. Massive CO and ONe white dwarfs ($m_{\rm wd}\gtrsim 0.6\,\msun)$, on the other hand, probably formed through a different channel of unstable Roche-lobe overflow during helium shell burning (Case C), followed by common envelope inspiral. The measured eccentricities of these massive white dwarfs remain to be explained.
\end{abstract}




\section{Introduction}

Thanks to their exquisite rotational stability, timing the radio pulses of binary millisecond pulsars (MSPs) provides extraordinarily precise measurements of their orbits. Specifically, MSPs with white dwarf (WD) companions exhibit some of the lowest eccentricities measured in astronomy, with values as low as $e\sim 10^{-7}$ \citep{PhinneyKukarni1994,Freire2012,EPTA2023,Chisabi2025}.\footnote{Another example is the detection of planetary-mass orbiting companions as light as the Moon \citep{Wolszczan1994}.}

With the exception of a few eccentric MSPs \citep[eMSPs; see, e.g.][and reference therein]{FreireTauris2014,HanLi2021,GinzburgChiang2022}, the low eccentricities of field MSPs with helium (He) WD companions are explained well by the convective fluctuation--dissipation theory of \citet{Phinney1992}. According to this theory, an MSP--He WD binary is formed through stable Roche-lobe overflow of the WD's red giant branch (RGB) low-mass progenitor ($M\lesssim 2\,\msun$), while it burns hydrogen (H) in a shell above its degenerate He core \citep[Case B overflow, or Case A for the shortest orbital periods, see][]{Paczynski1971,Tauris2011conf}. The strong tides that the pulsar raises on the convective envelope of the Roche-lobe filling giant dissipate the orbital eccentricity, but the same convection also perturbs the orbit and excites the eccentricity by stochastically fluctuating the giant's gravitational quadrupole moment. The residual eccentricity is determined by energy equipartition between the epicyclic motion and the convective eddies during Roche-lobe detachment, when the progenitor's remaining low-mass envelope contracts such that tides become ineffective. This process yields a distinct relation between the remnant He WD's orbital period and eccentricity, which fits the observations (impressively, the theory predicts not only the slope but also the normalization of this relation, such that there are no free parameters).

Interestingly, more massive carbon--oxygen (CO) WDs orbiting MSPs typically exhibit eccentricities in the same range $e\sim 10^{-6}-10^{-3}$ as their lower mass He counterparts \citep{Tauris2012,Hui2018,Parent2019}. Motivated by these similar eccentricities, \citet{Cohen2024} extended the \citet{Phinney1992} mechanism to asymptotic giant branch (AGB) stars that overflow their Roche lobe while burning He in a shell above their CO cores (Case C overflow), leaving behind MSP--CO WD binaries.
However, the analysis of \citet{Cohen2024} suffers from several shortcomings. First, in order to produce massive CO WDs through this channel, their progenitors might be too massive for stable overflow, leading to an inspiral within a common envelope \citep{Paczynski1976,IbenLivio1993,vanDenHeuvel994,Ivanova2013}. In fact, the observed orbits of CO WDs are orders of magnitude shorter than the Roche-lobe filling orbits of AGB stars, in stark contrast to He WDs, which fit the orbits of their RGB progenitors \citep{Rappaport1995,TaurisSavonije99}. Finally, \citet{Cohen2024} computed an approximately constant $e\sim 3\times 10^{-3}$, which bounds almost all of the observations from above, but cannot explain their span down to $e\sim 10^{-6}$. 

These shortcomings led \citet{Cohen2024} to consider an alternative formation path for MSP--CO WD binaries, which avoids the AGB and common envelope phases. For a certain region of the parameter space, intermediate-mass progenitors ($3\,\msun\lesssim M\lesssim 5\,\msun$) that overflow their Roche lobe during the main sequence (Case A) or during the RGB (Case B) transfer mass stably and burn their He cores into CO as they detach from the Roche lobe \citep{Tauris2000,Tauris2011conf,ShaoLi2012,Misra2020}.\footnote{See also \citet{KingRitter1999} and \citet{PodsiadlowskiRappaport2000}, who applied this scenario to the X-ray binary Cygnus X-2.} CO WDs formed through this channel reproduce the masses and orbital periods of a significant portion of the observations. In a preliminary calculation, \citet{Cohen2024} applied the \citet{Phinney1992} mechanism to a single typical test case, and demonstrated that the eccentricity in this channel $e\lesssim 10^{-4}$ is also representative of the observed population.

Here, we apply the convective fluctuation--dissipation theory \citep{Phinney1992} to the intermediate-mass stable Case A/B channel \citep{Tauris2000} systematically -- for an exhaustive range of initial progenitor masses and orbital periods. Our goal is to determine which of the observed MSP--CO WD binaries could have formed through this process. In addition to the the WD's final mass and orbital period, which have already been discussed in previous studies, we simultaneously compute -- for the first time -- the orbital eccentricity and compare it to the measurements. 

The remainder of this paper is organized as follows. In Section \ref{sec:radius}, we discuss the radius of the red giant progenitor. In Section \ref{sec:period}, we show how this radius determines the final orbital period. In Section \ref{sec:ecc}, we compute the eccentricity and compare it to the observations.
In each of Sections \ref{sec:radius}--\ref{sec:ecc} we first briefly repeat (for context) the results of previous studies for the low-mass progenitors of MSP--He WD binaries \citep{Phinney1992,Rappaport1995}, and then extend them to the intermediate-mass progenitors of MSP--CO WD binaries while emphasizing the differences. We summarize and discuss our results in Section \ref{sec:summary}.

\section{Red giant progenitor radius}\label{sec:radius}

\subsection{Low-mass progenitors}

A key ingredient in the formation of MSP--He WD binaries is the relation between the envelope radius $R$ and the growing He core mass $m_{\rm c}$ of their low-mass red giant progenitors. These progenitors harbour compact degenerate cores, whose radius $r_{\rm c}\propto m_{\rm c}^{-1/3}$ depends solely on $m_{\rm c}$.\footnote{Strictly speaking, the cores are not fully degenerate, such that thermal pressure at their outskirts leads to significant deviations from the cold $r_{\rm c}(m_{\rm c})$ relation, which we keep only for illustrative purposes \citep{RefsdalWeigert1970,MillerBertolami2022}.} As a result, the nuclear reaction rate in the thin burning H shell above the core, and the envelope's radius required to evacuate this luminosity are also determined by $m_{\rm c}$, almost irrespectively of the progenitor's total mass $M$. Specifically, \citet{Rappaport1995} find that $R\propto m_{\rm c}^{9/2}$, which can also be derived analytically under certain approximations \citep{RefsdalWeigert1970,Kippenhahn2012,GinzburgChiang2022}. The giant's radius must of course be larger than the radius during the main sequence $R_0$, such that
\begin{equation}\label{eq:radius_lowmass}
    R(m_{\rm c})\approx\max\left[R_0,R_0\left(\frac{m_{\rm c}}{m_0}\right)^{9/2}\right],
\end{equation}
where $m_0$ is the minimum core mass for which the star expands beyond its main sequence size (when the shell's luminosity exceeds the bulk H nuclear power).
For simplicity, we neglect the dependence of $R_0(M)$, such that $m_0\approx 0.2\,\msun$ is also independent of $M$.
We note that \citet{Rappaport1995} use a slightly different interpolation with the main sequence, and also include the limit $R\propto m_{\rm c}^{1/2}$ for the massive cores of AGB stars \citep[see][]{Cohen2024}, which are irrelevant for our current study. 

\subsection{Intermediate-mass progenitors}

Intermediate-mass red giants behave differently because their He cores are not degenerate. Let us now consider for simplicity an isothermal ideal gas He core at a temperature $T_{\rm c}$. From hydrostatic equilibrium, the core's pressure scales as
\begin{equation}\label{eq:p_core}
    p_{\rm c}\sim\frac{Gm_{\rm c}^2}{r_{\rm c}^4},
\end{equation}
where $G$ is the gravitational constant. On the other hand, the pressure is given by the ideal gas law 
\begin{equation}\label{eq:p_ideal}
    p_{\rm c}\sim\frac{\rho_{\rm c}}{\mu_{\rm c}m_{\rm u}}k_{\rm B}T_{\rm c},
\end{equation}
where $\rho_{\rm c}\sim m_{\rm c}r_{\rm c}^{-3}$ is the core's mass density, $\mu_{\rm c}$ is its (dimensionless) mean molecular weight, $m_{\rm u}$ is the atomic mass unit, and $k_{\rm B}$ is Boltzmann's constant. By comparing equations \eqref{eq:p_core} and \eqref{eq:p_ideal}, the core's radius is given by
\begin{equation}\label{eq:r_ideal}
    r_{\rm c}\sim\frac{Gm_{\rm c}\mu_{\rm c}m_{\rm u}}{k_{\rm B}T_{\rm c}}.
\end{equation}
We substitute $r_{\rm c}$ in either equation \eqref{eq:p_core} or \eqref{eq:p_ideal}, and find
\begin{equation}\label{eq:p_core_m}
    p_{\rm c}\sim\frac{k_{\rm B}^4}{G^3m_{\rm u}^4}\frac{T_{\rm c}^4}{m_{\rm c}^2\mu_{\rm c}^4}.
\end{equation}
We can apply the same calculation to the surrounding ideal gas envelope and estimate its pressure
\begin{equation}\label{eq:prs_e}
    p_{\rm e}\sim\frac{k_{\rm B}^4}{G^3m_{\rm u}^4}\frac{T_{\rm e}^4}{m_{\rm e}^2\mu_{\rm e}^4}\approx\frac{k_{\rm B}^4}{G^3m_{\rm u}^4}\frac{T_{\rm e}^4}{M^2\mu_{\rm e}^4},
\end{equation}
where $T_{\rm e}$, $m_{\rm e}$, and $\mu_{\rm e}$ are the envelope's temperature, mass, and mean molecular weight (not to be confused with the mean molecular weight per electron), respectively. We assume that $m_{\rm c}\ll M$, such that $m_{\rm e}=M-m_{\rm c}\approx M$. 

This core--envelope structure is stable only if $p_{\rm c}>p_{\rm e}$. Assuming thermal equilibrium between the core and the envelope $T_{\rm c}=T_{\rm e}$, the stability criterion implies that
\begin{equation}
    \frac{m_{\rm c}}{M}<\beta\approx 0.37\left(\frac{\mu_{\rm e}}{\mu_{\rm c}}\right)^2,
\end{equation}
where $\beta$ is the \citet{SC1942} limit. See \citet{Kippenhahn2012} for a more rigorous derivation, including the $\approx 0.37$ numerical coefficient, such that $\beta\approx 0.1$ for a pure He core ($\mu_{\rm c}=4/3$) and a solar composition envelope ($\mu_{\rm e}\approx 0.6$). Note that $\beta\ll 1$, justifying our $m_{\rm c}\ll M$ approximation.

At the end of the main sequence, the progenitor star develops an almost pure He core, which continues growing in mass through H burning in a shell surrounding the core. As long as $m_{\rm c}<\beta M$, the star remains close to its main sequence size. Once the \scl limit is exceeded, i.e. $m_{\rm c}>\beta M$ and therefore $p_{\rm c}<p_{\rm e}$, the core can no longer maintain thermal equilibrium with the envelope ($T_{\rm c}=T_{\rm e}$) while supporting against its pressure, such that the core contracts. Similarly to low-mass red giants, the contraction of the core is accompanied by the expansion of the envelope \citep[the `mirror principle'; e.g.][]{Kippenhahn2012}. However, in contrast to the degenerate cores of low-mass red giants, the ideal gas cores of intermediate-mass stars contract on a thermal time-scale, which is much shorter than the nuclear time-scale on which the core grows in mass \citep{Kippenhahn2012,MillerBertolami2022}. We parametrize the envelope's associated rapid expansion by adapting equation \eqref{eq:radius_lowmass}  
\begin{equation}\label{eq:rad_intermass}
    R(M,m_{\rm c})\approx\max\left[R_0,R_0\left(\frac{m_{\rm c}}{\beta M}\right)^\alpha\right],
\end{equation}
where $\alpha>9/2$ is a steep power, and with the expansion initiated at the \scl limit $m_{\rm c}=\beta M$. Depending on $M$, the progenitor star may exit the main sequence with a core mass $m_{\rm c}$ that already slightly exceeds the \scl limit, a product of its fully mixed convective core on the main sequence \citep{MillerBertolami2022}. We avoid this complication by parametrizing the onset of expansion with $\beta$, which in reality is the maximum of the \scl limit and the initial core mass fraction (in any case, by coincidence, the two values are similar for the relevant mass range). 

\begin{figure}
\includegraphics[width=\columnwidth]{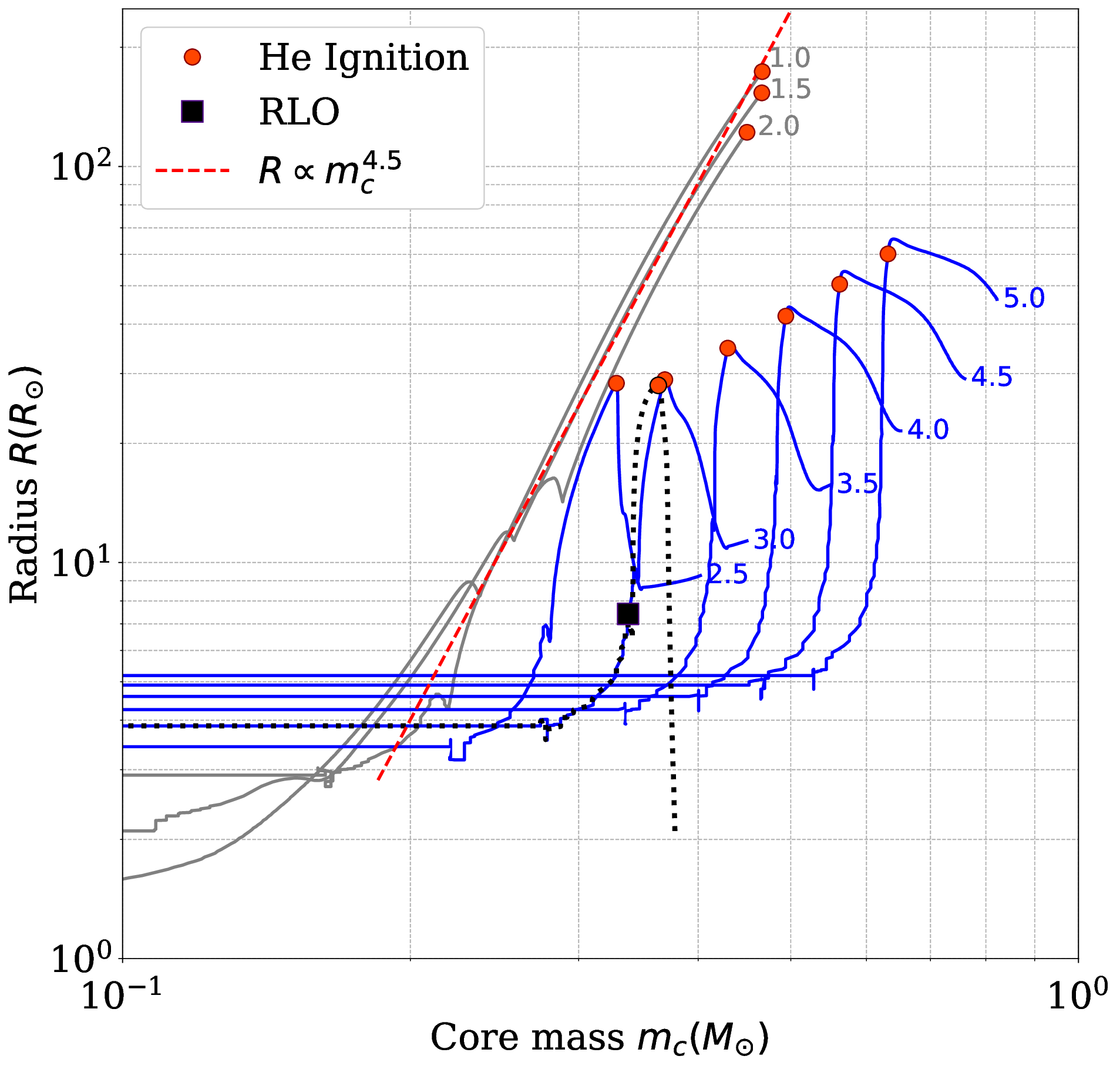}
\caption{Evolutionary tracks of low-mass (solid grey lines) and intermediate-mass (solid blue lines) single stars, with the stellar mass $M/\msun$ indicated near each track; the initial metallicity is $Z=0.02$. As stars exit the main sequence, they develop a He core with a growing mass $m_{\rm c}$ while their envelope's radius $R$ expands. Low-mass stars harbour degenerate cores and expand on almost the same $R\propto m_{\rm c}^{9/2}$ curve (dashed red line), given by equation \eqref{eq:radius_lowmass}. Intermediate-mass stars harbour non-degenerate cores and expand on different curves, given approximately by equation \eqref{eq:rad_intermass}, until the cores ignite (red circles) and the stars contract. Binary stars that undergo Roche-lobe overflow (RLO; black square) deviate from these single stellar evolutionary tracks and ignite He at a lower $m_{\rm c}$ (dotted black line; for an initially $3\,\msun$ star on a $4$ d orbit around an $M_{\rm ns}=1.8\,\msun$ neutron star).}
\label{fig:radius}
\end{figure}

As the He core contracts and heats up, it is eventually ignited, which results in a dramatic contraction of the envelope (see Fig. \ref{fig:radius}).\footnote{Intuitively, He burning in the core (which appears in addition to pre-existing H burning in a shell) sets a thermal and hydrostatic equilibrium that is more similar to a smaller He main sequence star compared to a purely shell-burning giant -- explaining why He ignition leads to stellar contraction \citep{Kippenhahn2012}.} Whereas low-mass progenitors (whose evolution is independent of their total mass $M$) ignite He in their degenerate cores at the same $m_{\rm c}\approx 0.45\,\msun$, intermediate-mass progenitors ignite He at different non-degenerate core masses $m_{\rm c}(M)$; see Fig. \ref{fig:radius}. As we demonstrate in Section \ref{sec:period}, this enables the creation of CO WDs with different masses (both above and below $0.45\,\msun$) in binary systems.

According to equation \eqref{eq:p_core_m}, for core masses at and below the \scl limit, the ideal gas pressure scales as $p_{\rm c}\propto m_{\rm c}^{-2}$, because $T_{\rm c}=T_{\rm e}$ is approximately constant: the envelope's temperature is essentially the H burning temperature of main sequence stars which varies very weakly with $M$. The degeneracy pressure, on the other hand, scales more steeply as
\begin{equation}\label{eq:p_deg}
    p_{\rm d}\propto \rho_{\rm c}^{5/3}\sim\left(\frac{m_{\rm c}}{r_{\rm c}^3}\right)^{5/3}\propto m_{\rm c}^{-10/3},
\end{equation}
where we have substituted $r_{\rm c}\propto m_{\rm c}$ from equation \eqref{eq:r_ideal}, such that sufficiently low-mass cores are actually degenerate (and our ideal gas treatment breaks down). Since at the \scl limit $M=m_{\rm c}/\beta\propto m_{\rm c}$, low-mass progenitors $M\lesssim m_0/\beta\approx2\,\msun$ harbour degenerate cores such that this limit becomes irrelevant, and $R$ follows equation \eqref{eq:radius_lowmass} instead of equation \eqref{eq:rad_intermass}.

In Fig. \ref{fig:radius} we use the stellar evolution code \textsc{mesa} \citep{Paxton2011,Paxton2013,Paxton2015,Paxton2018,Paxton2019,Jermyn2023}, version r24.08.1, to plot the radius $R$ during the evolution of low-mass ($M\lesssim 2\,\msun$) and intermediate-mass ($3\,\msun\lesssim M\lesssim 5\,\msun$) stars with an initial metallicity $Z=0.02$ (i.e. roughly solar). These single stellar evolutionary tracks were computed using the \texttt{star\_plus\_point\_mass} binary test suite with an effectively infinite initial orbital period (such that the star remains isolated throughout its evolution) for consistency with the binary stellar evolution calculations in Section \ref{sec:period}. As demonstrated in the figure, after the end of the main sequence, low-mass stars ascend the same RGB track, such that their radius $R(m_{\rm c})$ depends only on the mass of the core, almost independently of the envelope's mass and therefore of $M$.
Intermediate-mass stars, on the other hand, follow different tracks as a function of $M$, such that the radius $R(M,m_{\rm c})$ is a function of both the core mass and the total mass. This different behaviour -- the result of electron degeneracy in the core -- is captured, at least qualitatively, by equations \eqref{eq:radius_lowmass} and \eqref{eq:rad_intermass}. Quantitatively, we fit the parameters $R_0$, $\alpha$, and $\beta$ in Section \ref{sec:period} for a better agreement with the final orbital period (which is observable), rather than to the radius itself.

\section{Final orbital period}\label{sec:period}

In Section \ref{sec:radius} we described the evolution of single stars with mass $M$ as they ascend the RGB and expand. 
If, instead, such a star orbits a neutron star with a mass $M_{\rm ns}$, then, depending on the initial orbital period, it may overflow its Roche lobe and initiate mass transfer. If the mass transfer remains stable, the star continues to fill its Roche lobe such that 
\begin{equation}\label{eq:r_l}
    R=R_{\rm L}\approx 0.46 a\left(\frac{M}{M+M_{\rm ns}}\right)^{1/3},
\end{equation}
where $a$ is the orbital separation, and the Roche lobe's approximate radius $R_{\rm L}$ is given by \citet{Paczynski1971}.\footnote{We intentionally adopt this approximation instead of the more accurate \citet{Eggleton1983} fit (which is used in our \textsc{mesa} calculations) in order to derive simpler and more intuitive analytical expressions.}
From Kepler's laws and equation \eqref{eq:r_l}, the orbital period $P$ is given by
\begin{equation}\label{eq:period_general}
    P=2\upi\left[\frac{a^3}{G(M+M_{\rm ns})}\right]^{1/2}\approx 20 \left(\frac{R^3}{GM}\right)^{1/2},
\end{equation}
as in \citet{Rappaport1995}.

\subsection{Low-mass progenitors}\label{sec:period_lowmass}

A low-mass giant progenitor that overflows its Roche lobe follows equation \eqref{eq:period_general} until its H envelope's decreasing mass drops below a critical value $m_{\rm e}\sim 10^{-3}-10^{-2}\,\msun$ \citep[a few times the mass of the burning shell; see][]{RefsdalWeigert1969,RefsdalWeigert1971,Phinney1992,Cohen2024}. At this point, the star's radius rapidly contracts and it detaches from its Roche lobe -- ceasing the mass transfer and orbital evolution. Using equations \eqref{eq:radius_lowmass} and \eqref{eq:period_general}, and approximating $m_{\rm e}\ll m_{\rm c}\approx M$ at Roche-lobe detachment, we reproduce the relation $P\propto m_{\rm wd}^{25/4}$ between the final orbital period and the mass $m_{\rm c}$ of the He core, which forms the remnant He WD with mass $m_{\rm wd}\approx m_{\rm c}\lesssim 0.45\,\msun$. See \citet{Rappaport1995} for the same derivation and \citet{TaurisSavonije99} for a more accurate fit to numerical calculations \citep[see][for the original derivation]{RefsdalWeigert1971}. Specifically, \citet{TaurisSavonije99} find slightly shorter final orbital periods for WDs that originated from low-metallicity Population II stars.

\subsection{Intermediate-mass progenitors}\label{sec:period_intermass}

Intermediate-mass progenitors that stably overflow their Roche lobe (we discuss the stability below) also follow equation \eqref{eq:period_general}. However, their evolution is more complex because their radius $R(M,m_{\rm c})$ strongly depends on their total mass $M$, as well as on their core mass $m_{\rm c}$, as explained in Section \ref{sec:radius}. 
Moreover, while in Section \ref{sec:radius}  we analysed the evolutionary tracks of single stars that conserve their mass $M$, overflowing stars in binary systems gradually lose their envelope, such that $M$ decreases. 

As $M$ decreases, an overflowing star deviates from the single stellar evolutionary track that matches its initial mass $M$, and approaches the tracks of lower-mass stars (see Fig. \ref{fig:radius}). To understand this deviation intuitively, imagine a star that overflows its Roche lobe shortly after its core exceeds the \scl limit, such that there is a small pressure imbalance between the core and the envelope $p_{\rm e}>p_{\rm c}$. As $M$ decreases due to overflow, $p_{\rm e}$ increases further according to equation \eqref{eq:prs_e}, such that the core contracts (and therefore the envelope expands) to balance the pressures more than in the case of constant $M$. As can be appreciated from Fig. \ref{fig:radius}, this faster core contraction leads to He ignition at a smaller $m_{\rm c}$.
Similarly to single stars, He ignition in the core results in contraction of the envelope, detaching the star from its Roche lobe and thus setting its final mass $m_{\rm wd}$ and period $P$. The detached star then completes its He burning and forms a remnant CO WD.

Despite the complex evolution of overflowing intermediate-mass giants, we model the final orbital periods of their CO WD progenies using a simple adaptation of the \citet{Rappaport1995} analytical formula (i.e. Section \ref{sec:period_lowmass}). By combining equations \eqref{eq:rad_intermass} and \eqref{eq:period_general}, and substituting the final core mass $m_{\rm c}$ at Roche-lobe detachment (as in Section \ref{sec:period_lowmass}), we estimate 
\begin{equation}\label{eq:period_intermass}
    P\approx 20\left[\frac{R_0^3}{G(m_{\rm wd}-m_{\rm e})}\right]^{1/2}\max\left[1,\left(\frac{m_{\rm wd}-m_{\rm e}}{\beta M}\right)^{3\alpha/2}\right],
\end{equation}
where $M$ is the progenitor's \textit{initial} mass (which sets the \scl limit). In addition to replacing equation \eqref{eq:radius_lowmass} with equation \eqref{eq:rad_intermass}, another difference with respect to Section \ref{sec:period_lowmass} is explicitly substituting $m_{\rm wd}=m_{\rm c}+m_{\rm e}$ (instead of approximating $m_{\rm wd}\approx m_{\rm c}$), where $m_{\rm c}$ and $m_{\rm e}$ are the core and envelope masses at Roche-lobe detachment, which together comprise the final WD mass $m_{\rm wd}$. The motivation for this is twofold. First, due to the different detachment mechanism, which involves contraction at He ignition, $m_{\rm e}$ at Roche-lobe detachment is about an order of magnitude larger for intermediate-mass progenitors, so we can no longer neglect it. Second, the dependence $R\propto m_{\rm c}^\alpha$ on the core mass is steeper in equation \eqref{eq:rad_intermass}, $\alpha>9/2$ (see also the steeper slope in Fig. \ref{fig:radius}), such that the period at Roche-lobe detachment is more sensitive to the exact value of $m_{\rm c}$ at that moment. We note that there is an additional dependence on $m_{\rm c}^{-1/2}$ in the first term of equation \eqref{eq:period_intermass}, but it is much weaker so we can choose either $m_{\rm wd}$ or $m_{\rm c}$ for that term (we choose $m_{\rm c}$ for consistency and for a slightly better fit to the numerical curves).

Equation \eqref{eq:period_intermass} implies that a CO WD's final orbital period depends on its progenitor's initial mass $M$, in contrast to He WDs \citep[see also][]{Tauris2000}. A notable exception is stable Case C overflow of low-mass AGB stars \citep{Tauris2000,ShaoLi2012}; this channel was already covered by \citet{Cohen2024} -- here we focus instead on Case A/B overflow of intermediate-mass RGB stars. 

\addtocounter{footnote}{-1} 

\begin{figure}
\includegraphics[width=\columnwidth]{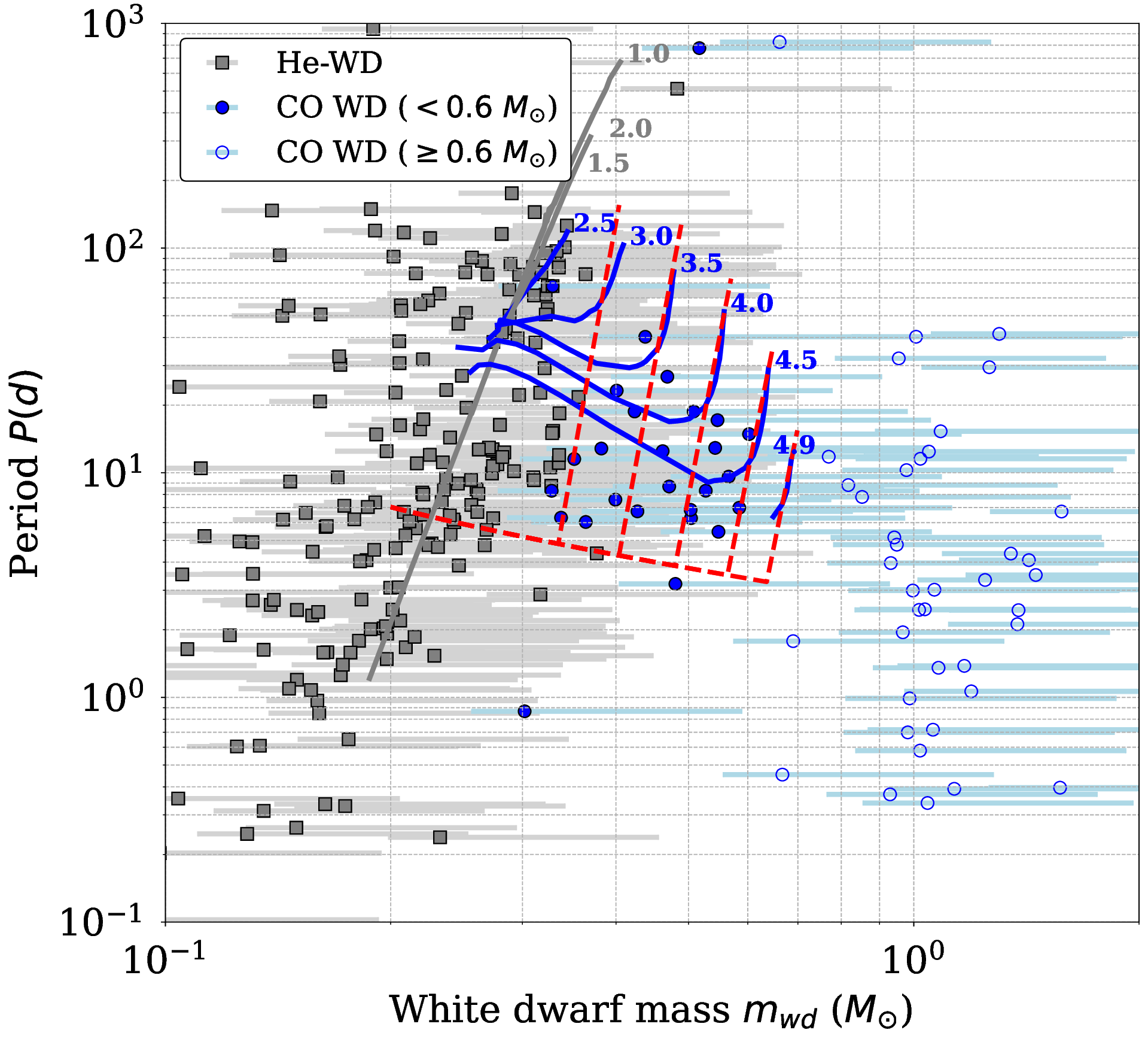}
\caption{Observed MSP--He WD (grey squares) and MSP--CO WD (filled and empty blue circles, these include ONe WDs) field binaries with pulsar spin periods below 50 ms from the ATNF Pulsar Catalogue \url{http://www.atnf.csiro.au/research/pulsar/psrcat} \citep{Manchester2005}, version 2.7.0 (October 2025). Markers represent median WD masses, and the error bars span from the minimum mass to the 90th percentile. Low-mass He WDs are explained by stable Roche-lobe overflow of low-mass progenitors (solid grey lines),\protect\footnotemark\,and intermediate-mass CO WDs (median $m_{\rm wd}<0.6\,\msun$, filled blue circles) are generally explained by stable Roche-lobe overflow of intermediate-mass progenitors (solid blue lines, with the initial progenitor mass $M/\msun$ indicated near each curve). A simple analytical model (dashed red lines), given by equation \eqref{eq:period_intermass}, approximately reproduces these numerical curves. Massive WDs (median $m_{\rm wd}>0.6\,\msun$, empty circles) probably formed through unstable overflow followed by common envelope inspiral.}
\label{fig:period}
\end{figure}

\footnotetext{The lowest-mass WDs $m_{\rm wd}\lesssim 0.2\,\msun$ were reproduced with a mixing length parameter $\alpha_{\rm mlt}=3$ instead of the nominal $\alpha_{\rm mlt}=2$. We note that both the theoretical and observational lower mass limit for He WDs is $m_{\rm wd}\approx 0.15\,\msun$ \citep{SunArras2018,Li2019,Brown2020}. The systems in Fig. \ref{fig:period} with median WD masses below this limit are likely observed at a low inclination angle (i.e. nearly face-on), such that their actual mass is considerably higher than the median (see the error bars).}

\begin{figure}
\includegraphics[width=\columnwidth]{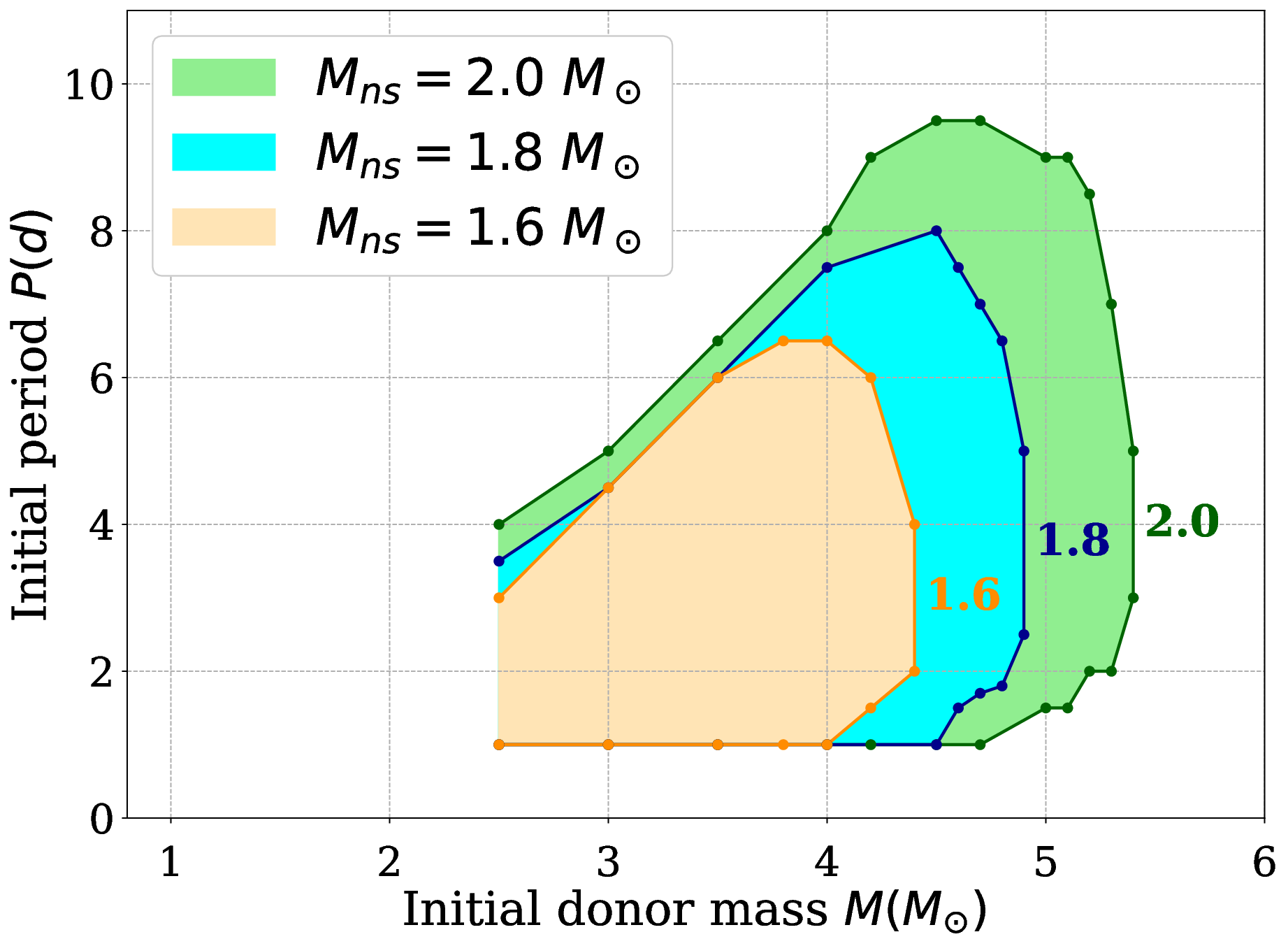}
\caption{The parameter space (initial donor mass $M$, initial neutron star mass $M_{\rm ns}$, and initial orbital period $P$) for stable Roche-lobe overflow of intermediate-mass progenitors (i.e. the channel we focus on in this study). Due to the donor to accretor mass ratio, low-mass progenitors ($M\lesssim 2\,\msun$) are always stable, whereas massive progenitors ($M\gtrsim 6\,\msun$) are always unstable. For intermediate masses, the stability depends on whether the star overflows its Roche lobe when its envelope is mostly radiative (short periods, stable) or convective (long periods, unstable). Outside this parameter space, unstable Roche-lobe overflow leads to common envelope inspiral.}
\label{fig:stability}
\end{figure}

In Fig. \ref{fig:period}, we plot the final orbital period as a function of the remnant WD's mass $P(m_{\rm wd})$ for intermediate-mass progenitors with initial masses $3\,\msun\lesssim M\lesssim 5\,\msun$ and different initial orbital periods, computed using the \texttt{star\_plus\_point\_mass} binary test suite of the \textsc{mesa} stellar evolution code. We choose a nominal initial neutron star (point) mass $M_{\rm ns}=1.8\,\msun$, which is relatively high but represents a considerable fraction of MSP--WD binaries \citep{Tauris2011,OzelFreire2016}. 
A low donor to accretor mass ratio tends to stabilize the mass transfer \citep[e.g.][]{Rappaport1982}, such that low-mass progenitors ($M\lesssim 2\,\msun$) always overflow their Roche lobe stably. More massive intermediate-mass progenitors, on the other hand, transfer mass stably only for a limited range of initial orbital periods, which is indicated in Fig. \ref{fig:stability} \citep[see also][]{Tauris2000,ShaoLi2012}. These orbital periods correspond to Roche-lobe overflow at the end of the main sequence (Case A) or beginning of H shell burning (early Case B), before the stellar envelope becomes mostly convective. 
The mostly radiative envelopes in this case contract in response to mass loss -- in contrast to convective envelopes -- stabilizing the mass transfer \citep{Tauris2000,Temmink2023}. 
As demonstrated in Fig. \ref{fig:stability}, the neutron star's mass determines (through the donor to accretor mass ratio) the maximal initial progenitor mass $M$ for which stable Roche-lobe overflow is possible (at least for some initial periods). The evolutionary trajectories and the region that they cover in the final $P-m_{\rm wd}$ plane change only mildly in the $M_{\rm ns}=1.6-2.0\,\msun$ range that we considered \citep[see also][]{ShaoLi2012}.

As seen in Fig. \ref{fig:period}, equation \eqref{eq:period_intermass} roughly reproduces the more accurate numerical calculations with the following choice of parameters: $R_0=3.5\,{\rm R}_{\sun}$, $\alpha=10$, $m_{\rm e}=0.08\,\msun$, and $\beta=0.1+0.015(M/\msun-4)$, such that it varies linearly between 0.085 and 0.11 in the stable mass range. This small deviation from a perfectly constant $\beta$ (e.g. the \scl limit $\beta\approx 0.1$) is necessary because $\beta$ is raised to a high power $3\alpha/2$. As seen in Fig. \ref{fig:menv}, our fitted $m_{\rm e}$ is an upper bound to the actual envelope mass shortly after Roche-lobe detachment, as extracted from our various \textsc{mesa} evolutionary tracks. Overall, our simple analytical model provides some valuable intuition with physically motivated parameters, but it deviates substantially at the shortest orbital periods from the more accurate numerical curves, which have the same morphology as in previous studies \citep{Tauris2000,ShaoLi2012,Nie2026}. 
The deviation of our analytical model at short orbital periods, which correspond to stars that hypothetically detach from their Roche lobe with $R\approx R_0$ in equation \eqref{eq:rad_intermass}, is not surprising. The cores of these main sequence stars are initially at the H burning temperature, such that they have to contract and heat up by an order unity factor in order to ignite He and thus detach from the Roche lobe. The contraction of the core is accompanied by an expansion of the envelope $R$, explaining the order of unity deviation of our analytical bottom limit from the numerical curves.

In Fig. \ref{fig:period} we also compare our theoretical curves to the observed MSP--CO WD population. We exclude MSPs in globular clusters because dynamical interactions with other stars in these dense environments could have affected their eccentricity \citep{RasioHeggie95}.
Regardless of any model, the currently observed sample is potentially bimodal, and can be separated according to the WD's mass $m_{\rm wd}$. The stable Case A/B channel considered in this study can explain mainly intermediate-mass CO WDs with $m_{\rm wd}\lesssim 0.6\,\msun$, whereas massive CO and potentially oxygen--neon (ONe) WDs with $m_{\rm wd}\gtrsim 0.6\,\msun$ probably formed through unstable Case C overflow followed by common envelope evolution \citep{Tauris2011conf,Cohen2024}. We note that our numerical curves do not reach the shortest period intermediate-mass CO WDs ($P\lesssim 10$ d). However, these systems have been reproduced by similar previous studies, e.g. with different neutron star masses, or by a small fraction of the transferred mass escaping through the L2 Lagrange point \citep{Tauris2000,Tauris2011,ShaoLi2012,Nie2026}.\footnote{The latter may be viewed more generally as one possible parametrization of the uncertainty in mass and angular momentum exchange in these binaries.} 
The couple of extremely long-period MSP--CO WDs in the sample \citep[$P\sim 10^3$ d; see][]{Wang2025} are likely products of stable Case C overflow of low-mass progenitors \citep{Tauris2000,ShaoLi2012,Cohen2024}, which is not the focus of this study. In any case, their eccentricity has not been measured yet.

\section{Eccentricity}\label{sec:ecc}

We are now poised in a position to estimate the eccentricity of MSP--WD binaries. According to \citet{Phinney1992}, the eccentricity is determined by energy equipartition between the epicyclic motion and the convective eddies in the red giant progenitor's envelope during Roche-lobe detachment \citep[see also][]{GinzburgChiang2022,Cohen2024}. Explicitly, the energy of the eccentric orbit (compared to a circular orbit with the same angular momentum) equals the kinetic energy of convective eddies
\begin{equation}\label{eq:equipartition}
    \frac{GM_{\rm ns}m_{\rm wd}}{2a}e^2\sim\frac{1}{2}m_{\rm e}v^2,
\end{equation}
where we have approximated $e\ll 1$. The convective velocity $v$ carries the giant's energy flux
\begin{equation}\label{eq:flux}
    F=\sigma T_{\rm eff}^4\sim\rho_{\rm e}v^3\sim\frac{m_{\rm e}v^3}{R^3},
\end{equation}
where $\sigma$ is the Stefan--Boltzmann constant, $\rho_{\rm e}$ is the envelope's mass density, and $T_{\rm eff}$ is the effective temperature. We follow \citet{Phinney1992} and take an approximately constant $T_{\rm eff}$ (because red giants follow the Hayashi line) and $M_{\rm ns}$, such that the eccentricity scales as
\begin{equation}\label{eq:ecc_step}
    e^2\propto am_{\rm wd}^{-1}m_{\rm e}v^2\propto am_{\rm wd}^{-1}m_{\rm e}^{1/3}R^2, 
\end{equation}
using equations \eqref{eq:equipartition} and \eqref{eq:flux}. Using Kepler's laws and equation \eqref{eq:r_l} at Roche-lobe detachment (when $M=m_{\rm wd})$, equation \eqref{eq:ecc_step} yields
\begin{equation}\label{eq:eP_scaling}
    e\propto P\left(\frac{m_{\rm e}}{m_{\rm wd}}\right)^{1/6},
\end{equation}
where we have also approximated $(1+m_{\rm wd}/M_{\rm ns})^{1/6}\approx 1$, which is accurate even for massive WDs.

\begin{figure}
\includegraphics[width=\columnwidth]{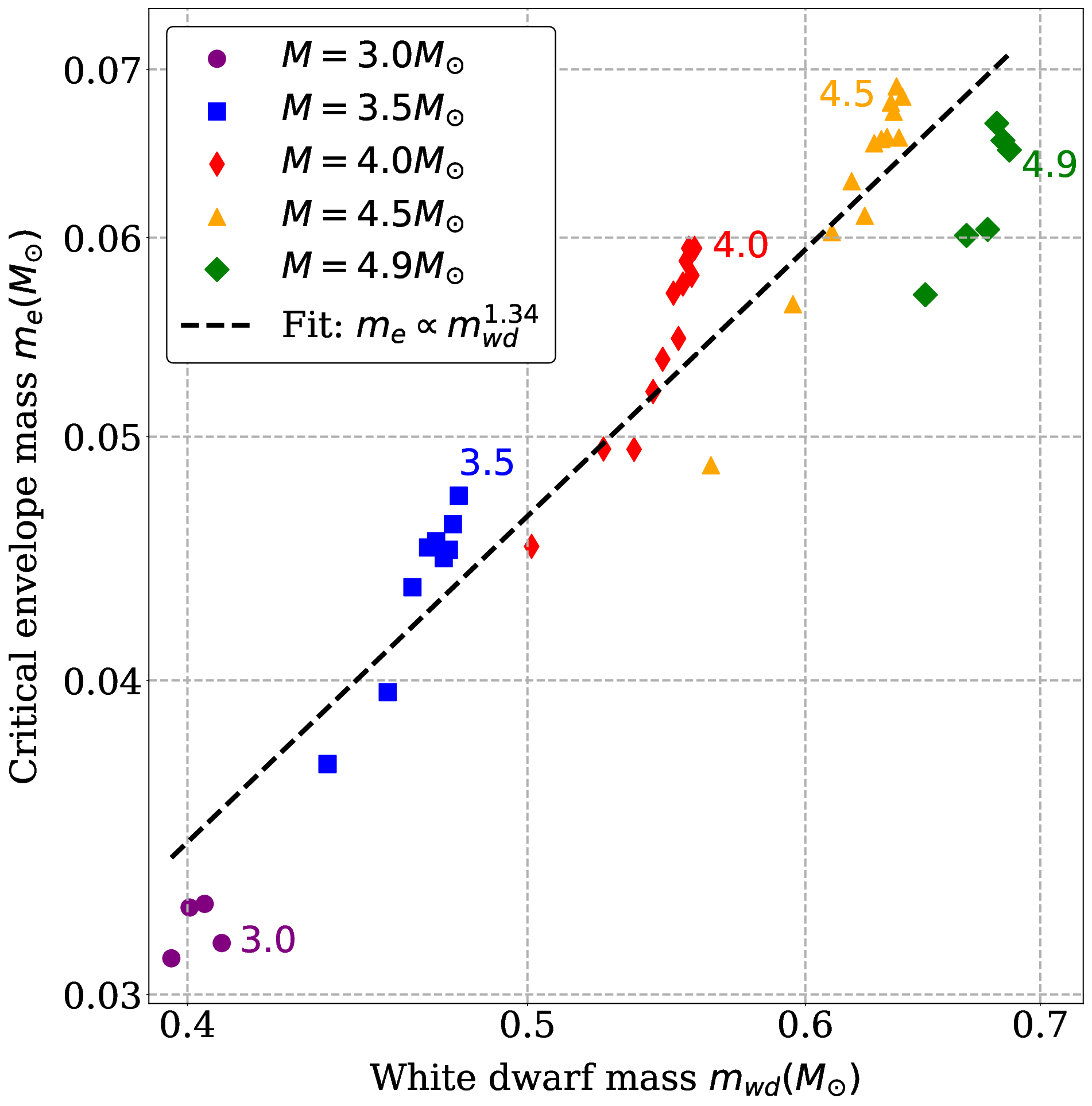}
\caption{The remaining envelope mass $m_{\rm e}$ (as a function of the initial progenitor mass $M$ and the final WD mass $m_{\rm wd}$) shortly after Roche-lobe detachment, when the envelope's radius contracts by a factor of 2 compared to the Roche lobe -- freezing the eccentricity \citep[we are not sensitive to the exact factor because of the sharp contraction; see][]{Cohen2024}. The intermediate-mass progenitors presented here ($3\,\msun\lesssim M\lesssim 5\,\msun$) detach from their Roche lobe upon He ignition, such that $m_{\rm e}$ is about an order of magnitude larger (and also scales differently with $m_{\rm wd}$) compared to low-mass progenitors that detach when their envelopes are too light to support a burning shell \citep{Cohen2024}.}
\label{fig:menv}
\end{figure}

More precisely, the final eccentricity is not determined at the moment of Roche-lobe detachment, but shortly afterwards. The time it takes convection to pump or damp the eccentricity is very sensitive to $a/R$, scaling as $\propto (a/R)^8$; see \citet{Zahn1977}. As long as the donor star fills its Roche lobe $a/R\approx 2.2 (1+M_{\rm ns}/M)^{1/3}$ according to equation \eqref{eq:r_l}, i.e. a factor of a few, and this time-scale is much shorter than the other evolutionary time-scales. However, once the star detaches and contracts by an additional factor of a few within its Roche lobe, this time becomes prohibitively long, freezing the eccentricity at a value set by equation \eqref{eq:eP_scaling} at that moment. In practice, $R$ drops sharply as a function of $m_{\rm e}$, such that $m_{\rm e}$ during the contraction is well defined \citep{Cohen2024}. We extract this $m_{\rm e}$ from our \textsc{mesa} calculations and plot it in Fig. \ref{fig:menv}.

\subsection{Low-mass progenitors}

\begin{figure}
\includegraphics[width=\columnwidth]{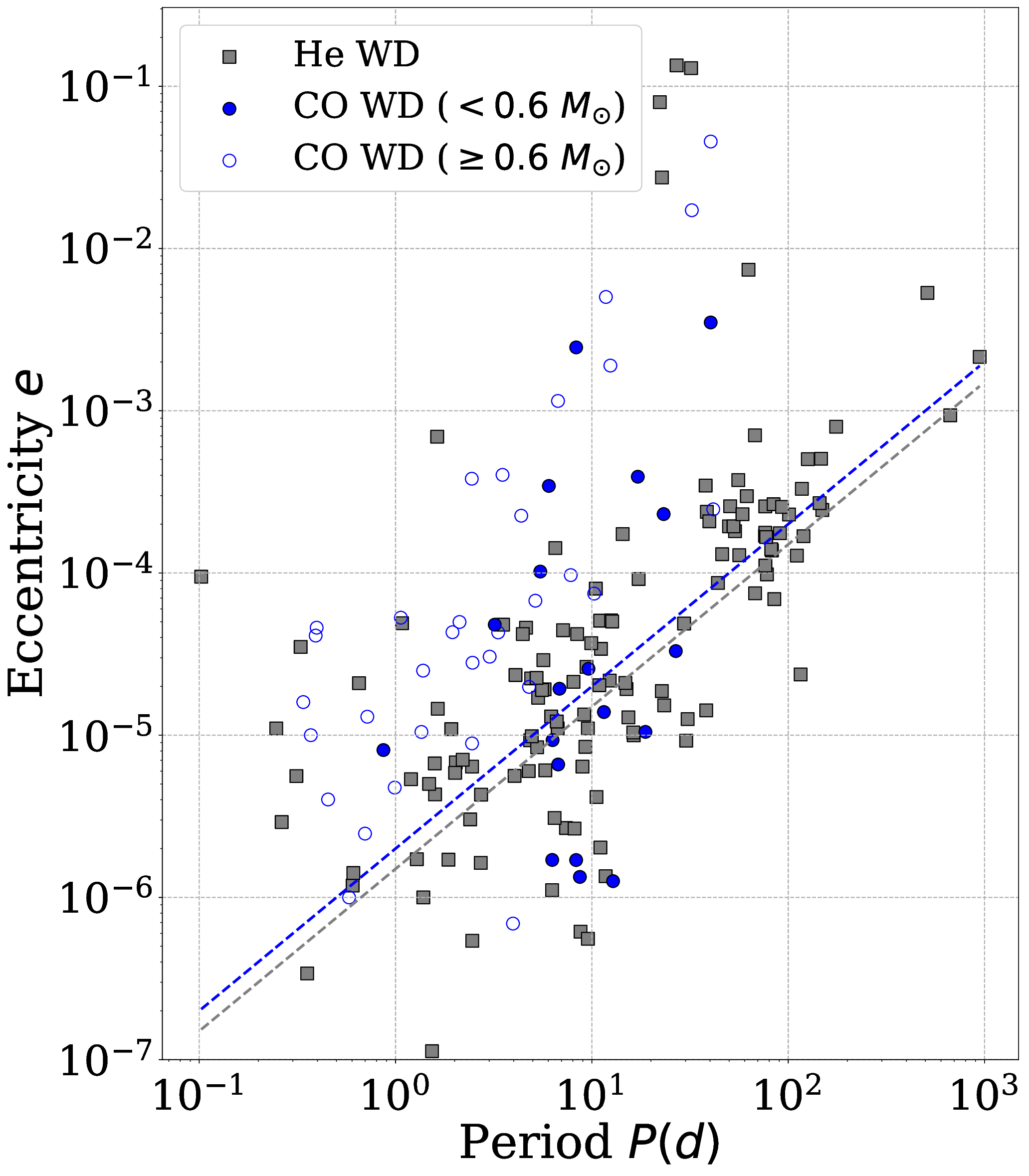}
\caption{Measured orbital periods $P$ and eccentricities $e$ of the observed binary MSP sample. Except for several eMSPs, low-mass He WD companions (grey squares) follow, on average, the \citet{Phinney1992} relation, given by equation \eqref{eq:ecc_low} -- the bottom dashed grey line. Intermediate-mass CO WDs (median $m_{\rm wd}<0.6\,\msun$, filled blue circles), which can be generally explained by stable Case A/B overflow (see Fig. \ref{fig:period}), follow a similar eccentricity distribution, but for a narrower range of periods, in agreement with equation \eqref{eq:ecc_inter} -- the top blue dashed line. High-mass CO and ONe WDs (median $m_{\rm wd}>0.6\,\msun$, empty circles) probably formed through a different channel of unstable Case C overflow and common envelope evolution -- resulting in a different eccentricity distribution, which has yet to be explained.} 
\label{fig:ecc}
\end{figure}

For low-mass progenitors, \citet{Phinney1992} approximated a constant $(m_{\rm e}/m_{\rm wd})^{1/6}$ in equation \eqref{eq:eP_scaling} by substituting $m_{\rm e}\approx 7\times 10^{-3}\,\msun$ (a few times the mass of the burning shell) for $m_{\rm wd}\approx 0.25\,\msun$ and neglecting the weak dependence of this term on $m_{\rm wd}$. After substituting all the other physical quantities, as well as including order of unity coefficients which we have omitted here, \citet{Phinney1992} derived the following normalized scaling: 
\begin{equation}\label{eq:ecc_low}
    e\approx 1.5\times 10^{-5}\left(\frac{P}{10\,\textrm{d}}\right),
\end{equation}
which fits well the average eccentricity of MSP--He WD binaries (excluding the eMSPs; see Fig. \ref{fig:ecc}).
\citet{Cohen2024} justified this approximation by extracting $m_{\rm e}\approx 7\times 10^{-3}\,\msun (m_{\rm wd}/0.25\,\msun)^{-5/2}$ from \textsc{mesa} computations, such that $(m_{\rm e}/m_{\rm wd})^{1/6}\propto m_{\rm wd}^{-7/12}\propto P^{-7/75}$ (see Section \ref{sec:period_lowmass} for the last scaling), which can be safely neglected in comparison to the $e\propto P$ term in equation \eqref{eq:eP_scaling}.

\subsection{Intermediate-mass progenitors}

When extending the \citet{Phinney1992} mechanism to intermediate-mass progenitors, all the steps in the derivation of equation \eqref{eq:eP_scaling} are still applicable. The only difference is the envelope's mass $m_{\rm e}$, which is about an order of magnitude higher in this case (Fig. \ref{fig:menv}), because the progenitor detaches from its Roche lobe upon core He ignition, regardless of the burning H shell.
None the less, thanks to the weak dependence $e\propto m_{\rm e}^{1/6}$ in equation \eqref{eq:eP_scaling}, this difference changes the normalization by only about $10^{1/6}\approx 1.5$, such that 

\begin{equation}\label{eq:ecc_inter}
    e\approx 2\times 10^{-5}\left(\frac{P}{10\,\textrm{d}}\right),
\end{equation}
which for all purposes is identical to the original normalization, given the stochastic nature of convection and the scatter in the observed eccentricities. 

Critically, equation \eqref{eq:ecc_inter} predicts eccentricities that are lower by two orders of magnitude than \citet{Cohen2024}, simply because in the Case A/B channel considered here the period at Roche-lobe detachment $P\sim 10$ d is much shorter than $P\sim 10^3$ d in the Case C scenario considered there. \citet{Cohen2024} invoked a subsequent phase of common envelope inspiral to reconcile such long \textit{initial} periods during Roche-lobe overflow with the \textit{final} observed $P\sim 10$ d. As seen in Fig. \ref{fig:period}, our stable Case A/B scenario produces intermediate-mass CO WDs ($m_{\rm wd}\lesssim 0.6\,\msun$) at the observed periods, without any common envelope phase.
  
As seen in Fig. \ref{fig:ecc}, apart from clustering in a narrower range of orbital periods (see also Fig. \ref{fig:period}), the period--eccentricity distribution of these intermediate-mass MSP--CO WDs is similar to that of low-mass MSP--He WDs, in agreement with equations \eqref{eq:ecc_low} and \eqref{eq:ecc_inter}. Quantitatively, in the period range $1\textrm{ d}<P<100\textrm{ d}$ where both groups overlap, intermediate-mass CO WDs have a $\log e=-4.4\pm 1.1$ (i.e. average $\pm$ one standard deviation), compared to $\log e=-4.7\pm 0.8$ ($\log e=-4.5\pm 1.1$) for He WDs with the eMSPs excluded (included). Similar results are obtained when considering narrower ranges of $3\textrm{ d}<P<30\textrm{ d}$ or $5\textrm{ d}<P<20\textrm{ d}$, i.e. the eccentricities of both WD groups spread about an order of magnitude in each direction even at the same orbital period. We emphasize that similarly to the original \citet{Phinney1992} scaling, equation \eqref{eq:ecc_inter} has no free parameters, i.e. both its slope and normalization are derived theoretically rather than fitted to the observations. High-mass ($m_{\rm wd}\gtrsim 0.6\,\msun$) CO and ONe WDs orbiting MSPs likely formed through Case C overflow followed by common envelope evolution, and therefore they follow a different eccentricity distribution \citep{Tauris2011conf,Cohen2024}.

\section{Summary and discussion}\label{sec:summary}

The formation history of MSP--He WD binaries has been well established for decades. Case B stable Roche-lobe overflow of low-mass ($M\lesssim 2\,\msun$) RGB progenitors (and for the shortest orbital periods, Case A overflow of main sequence stars) reproduces the masses, orbital periods, and eccentricities of these systems \citep{RefsdalWeigert1971,Joss1987,Savonije1987,Phinney1992,PhinneyKukarni1994,Rappaport1995,TaurisSavonije99}. The formation of MSPs with heavier CO WD companions is more complicated, with several possible formation channels \citep{Tauris2011conf,Tauris2011,Tauris2012}. 
Despite this difference, the eccentricities of both WD types overlap, with the vast majority of observations spanning the range $e\sim 10^{-6}-10^{-3}$ \citep{Tauris2012}. However, whereas the eccentricities of MSP--He WD binaries are explained well by the convective fluctuation--dissipation theory which relates them to the orbital period $e\propto P$ \citep{Phinney1992}, the eccentricities of MSP--CO WD binaries have so far been mostly fitted empirically -- without an underlying theory \citep{Hui2018,Parent2019}. 

In an initial attempt to theoretically explain MSP--CO WD eccentricities, \citet{Cohen2024} extended the \citet{Phinney1992} mechanism from low-mass RGB stars that overflow their Roche lobe while burning H in a shell above their degenerate He cores (Case B), to stars that overflow their Roche lobe later -- when they ascend the AGB and burn He in a shell above their CO cores (Case C). However, this scenario is not directly applicable to the vast majority of observed systems, because their short orbital periods require an additional phase of common envelope inspiral in this case. 

Here, we extended the \citet{Phinney1992} mechanism to a different MSP--CO WD formation channel. For a limited range of initial orbital periods, intermediate-mass progenitors ($3\,\msun\lesssim M\lesssim 5\,\msun$) overflow their Roche lobe at the end of the main sequence (Case A) or the beginning of the RGB (Case B), before their envelopes become mostly convective, ensuring the mass transfer's stability despite the initially high donor to accretor mass ratio. During the mass transfer, the non-degenerate He cores of these progenitors contract and ignite -- eventually leaving behind CO WDs without evolving through a common envelope phase \citep{Tauris2000,ShaoLi2012}. 

We used \textsc{mesa} to compute binary stellar evolution tracks covering the range of initial progenitor masses and orbital periods that yield MSP--CO WD binaries through this stable Case A/B overflow channel. We supplemented our numerical calculations \citep[which reproduce previous studies by][]{Tauris2000,ShaoLi2012} with a simple analytical formula $P(M,m_{\rm wd})$ linking the CO WD's final orbital period $P$ to its final mass $m_{\rm wd}$ and to its progenitor's initial mass $M$. We derived this formula by adapting the \citet{Rappaport1995} $P(m_{\rm wd})$ model from the degenerate He cores of low-mass progenitors to the non-degenerate He cores of intermediate-mass stars. With several physically motivated parameters (such as the radius on the main sequence $R_0\approx 3.5\,{\rm R}_{\sun}$, the \scl limit $\beta\approx 0.1$, and the envelope's mass at Roche-lobe detachment $m_{\rm e}\approx 0.08\,\msun$), our analytical equation \eqref{eq:period_intermass} reasonably fits the more accurate numerical curves, while providing physical intuition (see Fig. \ref{fig:period}).\footnote{To place these fitted parameters in context, stars in the relevant mass range have main-sequence radii $R_0\approx 3-5\,{\rm R}_{\sun}$ (Fig. \ref{fig:radius}), detach with envelope masses $m_{\rm e}\approx 0.03-0.07\,\msun$ (Fig. \ref{fig:menv}), and our fitted range for $\beta$ is $0.085-0.11$ (Section \ref{sec:period_intermass}).} 

We compared our theoretical curves in the $P-m_{\rm wd}$ plane to the currently observed sample of MSP--CO WD binaries in Fig. \ref{fig:period}. The observed population is bimodal and may be separated into intermediate mass CO WDs ($m_{\rm wd}\lesssim 0.6\,\msun$) and massive CO and ONe WDs ($m_{\rm wd}\gtrsim 0.6\,\msun$). We found that the stable Case A/B channel can explain primarily the intermediate-mass CO WDs, whereas the massive WDs probably formed through unstable Case C overflow followed by common envelope evolution \citep[see also][]{Tauris2011conf}.  

We applied the \citet{Phinney1992} mechanism to MSP--CO WD binaries that form through the stable Case A/B channel. The main difference compared to MSP--He WDs is the remaining H envelope mass shortly after Roche-lobe detachment $m_{\rm e}$, when the envelope contracts drastically. The convective eddies in this envelope set the eccentricity by stochastically fluctuating the star's gravitational quadrupole moment \citep{Phinney1992,GinzburgChiang2022}. We extracted $m_{\rm e}$ from our \textsc{mesa} computations, and found that it is higher by about an order of magnitude compared to He WDs \citep{Cohen2024}. The reason is that
the low-mass progenitors of He WDs detach from the Roche lobe when their envelopes become too light to support a burning H shell, whereas intermediate-mass progenitors of CO WDs detach when they ignite He in their cores. None the less, due to the weak dependence $e\propto m_{\rm e}^{1/6}$, this difference has only a minor effect on the eccentricity. We conclude that the original \citet{Phinney1992} scaling extends to intermediate-mass CO WDs, with only a slight change in normalization compared to lower mass He WDs.  

We divided the observed MSP--WD population into three groups in the $P-e$ plane (Fig. \ref{fig:ecc}): low-mass He WDs, intermediate-mass CO WDs that can be explained by stable Roche-lobe overflow (median $m_{\rm wd}<0.6\,\msun$), and massive CO and ONe WDs that probably experienced a common envelope phase (median $m_{\rm wd}>0.6\,\msun$). As we theoretically expected, $e(P)$ of the the first two groups follows a very similar distribution \citep[which is consistent with][]{Phinney1992}, with the only difference being the narrower period range of intermediate-mass CO WDs, presumably a result of their fine-tuned formation scenario (see Fig. \ref{fig:period}). Massive WDs in the third group generally have higher eccentricities at the same orbital period, as already found by \citet{Cohen2024}. We note that due to the large scatter in $e$ and the narrow range of orbital periods $P$, it is difficult to test the power law $e\propto P$ for intermediate-mass CO WDs. In fact, \citet{Hui2018} empirically fitted a much steeper, but low-significance, $e\propto P^4$ correlation when including the massive WDs as well. However, we emphasize that the convective fluctuation--dissipation theory provides not only a correlation, but also a concrete normalization. Specifically, we expect $e\sim 10^{-5}$ for the typical $P\sim 10$ d periods of intermediate-mass CO WDs, which is consistent with the observations (notwithstanding the scatter around this average value, which is similar for He WDs and intermediate-mass CO WDs).

Interestingly, the large scatter in the eccentricity has not received enough attention in the literature. \citet{Phinney1992} argued that $e^2$, which is proportional to the energy as seen in equation \eqref{eq:equipartition}, should be distributed according to the Boltzmann distribution. In this case, the eccentricity itself $e$ at a fixed orbital period $P$ should vary by a factor of a few around its mean value. As discussed in Section \ref{sec:ecc}, the observed standard deviation is somewhat larger -- about an order of magnitude in each direction for both He WDs and intermediate-mass CO WDs (for a fixed period $P\approx 10\textrm{ d})$. Such a large span potentially indicates a variation in the system's formation process or in its progenitor's properties, in addition to the stochastic convective scatter that \citet{Phinney1992} considered. We plan to investigate the observational and theoretical scatter more thoroughly in future work.

In summary, \citet{Tauris2000} first demonstrated that stable Case A/B Roche-lobe overflow of intermediate-mass progenitors reproduces the masses and orbital periods of observed intermediate-mass CO WDs orbiting MSPs. Here, we confirmed -- for the first time -- that the measured eccentricities of these systems are also reproduced by this process, similarly to the \citet{Phinney1992} mechanism for lower-mass He WDs -- firmly establishing this formation channel. According to this interpretation, the most massive CO and ONe WDs form a distinct group, whose eccentricity is (at least partially) determined by the uncertain physics of common envelope evolution instead \citep{GlanzPerets2021,Szolgyen2022,Trani2022}. Theoretically explaining the measured eccentricities of these massive WDs remains a challenge for future studies. 
  
\section*{Acknowledgements}

We thank Ravid Achituv, Omer Amsel, Omri Konstantino, and Shahar Nadav for stimulating discussions and useful comments. We also thank the anonymous reviewer for a detailed and constructive report which improved the paper.
This research was partially supported by the United States-Israel Binational Science Foundation (BSF; grant no. 2022175), the German-Israeli Foundation for Scientific Research and Development (GIF; grant no. I-1567-303.5-2024), and the Israel Science Foundation (ISF; grant nos 1600/24 and 1965/24).

\section*{Data Availability}
 
The data underlying this article will be shared on reasonable request to the corresponding authors.



\bibliographystyle{mnras}
\bibliography{pulsars} 

@ARTICLE{Phinney1992,
       author = {{Phinney}, E.~S.},
        title = "{Pulsars as Probes of Newtonian Dynamical Systems}",
      journal = {Philosophical Transactions of the Royal Society of London Series A},
         year = 1992,
        month = oct,
       volume = {341},
       number = {1660},
        pages = {39-75},
          doi = {10.1098/rsta.1992.0084},
       adsurl = {https://ui.adsabs.harvard.edu/abs/1992RSPTA.341...39P}
}

@ARTICLE{PhinneyKukarni1994,
       author = {{Phinney}, E.~S. and {Kulkarni}, S.~R.},
        title = "{Binary and Millisecond Pulsars}",
      journal = {\araa},
         year = 1994,
        month = jan,
       volume = {32},
        pages = {591-639},
          doi = {10.1146/annurev.aa.32.090194.003111},
       adsurl = {https://ui.adsabs.harvard.edu/abs/1994ARA&A..32..591P}
}

@ARTICLE{Tauris2000,
       author = {{Tauris}, Thomas M. and {van den Heuvel}, Edward P.~J. and {Savonije}, Gerrit J.},
        title = "{Formation of Millisecond Pulsars with Heavy White Dwarf Companions:Extreme Mass Transfer on Subthermal Timescales}",
      journal = {\apjl},
         year = 2000,
        month = feb,
       volume = {530},
       number = {2},
        pages = {L93-L96},
          doi = {10.1086/312496},
archivePrefix = {arXiv},
       eprint = {astro-ph/0001013},
 primaryClass = {astro-ph},
       adsurl = {https://ui.adsabs.harvard.edu/abs/2000ApJ...530L..93T}
}

@ARTICLE{ShaoLi2012,
       author = {{Shao}, Yong and {Li}, Xiang-Dong},
        title = "{Formation of Millisecond Pulsars from Intermediate- and Low-Mass X-Ray Binaries}",
      journal = {\apj},
         year = 2012,
        month = sep,
       volume = {756},
       number = {1},
          eid = {85},
        pages = {85},
          doi = {10.1088/0004-637X/756/1/85},
archivePrefix = {arXiv},
       eprint = {1207.2833},
 primaryClass = {astro-ph.HE},
       adsurl = {https://ui.adsabs.harvard.edu/abs/2012ApJ...756...85S}
}

@ARTICLE{Misra2020,
       author = {{Misra}, D. and {Fragos}, T. and {Tauris}, T.~M. and {Zapartas}, E. and {Aguilera-Dena}, D.~R.},
        title = "{The origin of pulsating ultra-luminous X-ray sources: Low- and intermediate-mass X-ray binaries containing neutron star accretors}",
      journal = {\aap},
         year = 2020,
        month = oct,
       volume = {642},
          eid = {A174},
        pages = {A174},
          doi = {10.1051/0004-6361/202038070},
archivePrefix = {arXiv},
       eprint = {2004.01205},
 primaryClass = {astro-ph.HE},
       adsurl = {https://ui.adsabs.harvard.edu/abs/2020A&A...642A.174M}
}

@ARTICLE{Cohen2024,
       author = {{Cohen}, Yair and {Ginzburg}, Sivan and {Levy}, Maya and {Bar Shalom}, Tal and {Siman Tov}, Yoav},
        title = "{White dwarf eccentricity fluctuation and dissipation by AGB convection}",
      journal = {\mnras},
         year = 2024,
        month = oct,
       volume = {534},
       number = {1},
        pages = {455-464},
          doi = {10.1093/mnras/stae2136},
archivePrefix = {arXiv},
       eprint = {2405.03745},
 primaryClass = {astro-ph.SR},
       adsurl = {https://ui.adsabs.harvard.edu/abs/2024MNRAS.534..455C}
}

@ARTICLE{EPTA2023,
       author = {{EPTA Collaboration} and {Antoniadis}, J. and {Babak}, S. and {Bak Nielsen}, A.-S. and {Bassa}, C.~G. and {Berthereau}, A. and {Bonetti}, M. and {Bortolas}, E. and {Brook}, P.~R. and {Burgay}, M. and {Caballero}, R.~N. and {Chalumeau}, A. and {Champion}, D.~J. and {Chanlaridis}, S. and {Chen}, S. and {Cognard}, I. and {Desvignes}, G. and {Falxa}, M. and {Ferdman}, R.~D. and {Franchini}, A. and {Gair}, J.~R. and {Goncharov}, B. and {Graikou}, E. and {Grie{\ss}meier}, J.-M. and {Guillemot}, L. and {Guo}, Y.~J. and {Hu}, H. and {Iraci}, F. and {Izquierdo-Villalba}, D. and {Jang}, J. and {Jawor}, J. and {Janssen}, G.~H. and {Jessner}, A. and {Karuppusamy}, R. and {Keane}, E.~F. and {Keith}, M.~J. and {Kramer}, M. and {Krishnakumar}, M.~A. and {Lackeos}, K. and {Lee}, K.~J. and {Liu}, K. and {Liu}, Y. and {Lyne}, A.~G. and {McKee}, J.~W. and {Main}, R.~A. and {Mickaliger}, M.~B. and {Ni{\c{t}}u}, I.~C. and {Parthasarathy}, A. and {Perera}, B.~B.~P. and {Perrodin}, D. and {Petiteau}, A. and {Porayko}, N.~K. and {Possenti}, A. and {Quelquejay Leclere}, H. and {Samajdar}, A. and {Sanidas}, S.~A. and {Sesana}, A. and {Shaifullah}, G. and {Speri}, L. and {Spiewak}, R. and {Stappers}, B.~W. and {Susarla}, S.~C. and {Theureau}, G. and {Tiburzi}, C. and {van der Wateren}, E. and {Vecchio}, A. and {Venkatraman Krishnan}, V. and {Verbiest}, J.~P.~W. and {Wang}, J. and {Wang}, L. and {Wu}, Z.},
        title = "{The second data release from the European Pulsar Timing Array. I. The dataset and timing analysis}",
      journal = {\aap},
         year = 2023,
        month = oct,
       volume = {678},
          eid = {A48},
        pages = {A48},
          doi = {10.1051/0004-6361/202346841},
archivePrefix = {arXiv},
       eprint = {2306.16224},
 primaryClass = {astro-ph.HE},
       adsurl = {https://ui.adsabs.harvard.edu/abs/2023A&A...678A..48E}
}

@ARTICLE{Chisabi2025,
       author = {{Chisabi}, M. and {Andrianomena}, S. and {Enwelum}, U. and {Gasennelwe}, E.~G. and {Idris}, A. and {Idogbe}, E.~A. and {Shilunga}, S. and {Geyer}, M. and {Reardon}, D.~J. and {Okany}, C.~F. and {Shamohammadi}, M. and {Shannon}, R.~M. and {Krishnan}, V. Venkatraman and {Abbate}, F. and {Kramer}, M.},
        title = "{Timing and noise analysis of five millisecond pulsars observed with MeerKAT}",
      journal = {\mnras},
         year = 2025,
        month = mar,
       volume = {537},
       number = {3},
        pages = {2462-2470},
          doi = {10.1093/mnras/staf100},
archivePrefix = {arXiv},
       eprint = {2501.07728},
 primaryClass = {astro-ph.HE},
       adsurl = {https://ui.adsabs.harvard.edu/abs/2025MNRAS.537.2462C}
}

@ARTICLE{Freire2012,
       author = {{Freire}, Paulo C.~C. and {Wex}, Norbert and {Esposito-Far{\`e}se}, Gilles and {Verbiest}, Joris P.~W. and {Bailes}, Matthew and {Jacoby}, Bryan A. and {Kramer}, Michael and {Stairs}, Ingrid H. and {Antoniadis}, John and {Janssen}, Gemma H.},
        title = "{The relativistic pulsar-white dwarf binary PSR J1738+0333 - II. The most stringent test of scalar-tensor gravity}",
      journal = {\mnras},
         year = 2012,
        month = jul,
       volume = {423},
       number = {4},
        pages = {3328-3343},
          doi = {10.1111/j.1365-2966.2012.21253.x},
archivePrefix = {arXiv},
       eprint = {1205.1450},
 primaryClass = {astro-ph.GA},
       adsurl = {https://ui.adsabs.harvard.edu/abs/2012MNRAS.423.3328F}
}

@ARTICLE{Wolszczan1994,
       author = {{Wolszczan}, Alexander},
        title = "{Confirmation of Earth-Mass Planets Orbiting the Millisecond Pulsar PSR B1257+12}",
      journal = {Science},
         year = 1994,
        month = apr,
       volume = {264},
       number = {5158},
        pages = {538-542},
          doi = {10.1126/science.264.5158.538},
       adsurl = {https://ui.adsabs.harvard.edu/abs/1994Sci...264..538W}
}

@ARTICLE{GinzburgChiang2022,
       author = {{Ginzburg}, Sivan and {Chiang}, Eugene},
        title = "{Eccentric millisecond pulsars by resonant convection}",
      journal = {\mnras},
         year = 2022,
        month = jan,
       volume = {509},
       number = {1},
        pages = {L1-L5},
          doi = {10.1093/mnrasl/slab110},
archivePrefix = {arXiv},
       eprint = {2109.10361},
 primaryClass = {astro-ph.SR},
       adsurl = {https://ui.adsabs.harvard.edu/abs/2022MNRAS.509L...1G}
}

@ARTICLE{FreireTauris2014,
       author = {{Freire}, Paulo C.~C. and {Tauris}, Thomas M.},
        title = "{Direct formation of millisecond pulsars from rotationally delayed accretion-induced collapse of massive white dwarfs}",
      journal = {\mnras},
         year = 2014,
        month = feb,
       volume = {438},
       number = {1},
        pages = {L86-L90},
          doi = {10.1093/mnrasl/slt164},
archivePrefix = {arXiv},
       eprint = {1311.3478},
 primaryClass = {astro-ph.SR},
       adsurl = {https://ui.adsabs.harvard.edu/abs/2014MNRAS.438L..86F}
}

@ARTICLE{HanLi2021,
       author = {{Han}, Qin and {Li}, Xiang-Dong},
        title = "{Asymmetrical Mass Ejection from Proto-white Dwarfs and the Formation of Eccentric Millisecond Pulsar Binaries}",
      journal = {\apj},
         year = 2021,
        month = mar,
       volume = {909},
       number = {2},
          eid = {161},
        pages = {161},
          doi = {10.3847/1538-4357/abdd21},
archivePrefix = {arXiv},
       eprint = {2101.12433},
 primaryClass = {astro-ph.HE},
       adsurl = {https://ui.adsabs.harvard.edu/abs/2021ApJ...909..161H}
}

@ARTICLE{Hui2018,
       author = {{Hui}, C.~Y. and {Wu}, Kinwah and {Han}, Qin and {Kong}, A.~K.~H. and {Tam}, P.~H.~T.},
        title = "{On the Orbital Properties of Millisecond Pulsar Binaries}",
      journal = {\apj},
         year = 2018,
        month = sep,
       volume = {864},
       number = {1},
          eid = {30},
        pages = {30},
          doi = {10.3847/1538-4357/aad5ec},
archivePrefix = {arXiv},
       eprint = {1807.09001},
 primaryClass = {astro-ph.HE},
       adsurl = {https://ui.adsabs.harvard.edu/abs/2018ApJ...864...30H}
}

@ARTICLE{Parent2019,
       author = {{Parent}, E. and {Kaspi}, V.~M. and {Ransom}, S.~M. and {Freire}, P.~C.~C. and {Brazier}, A. and {Camilo}, F. and {Chatterjee}, S. and {Cordes}, J.~M. and {Crawford}, F. and {Deneva}, J.~S. and {Ferdman}, R.~D. and {Hessels}, J.~W.~T. and {van Leeuwen}, J. and {Lyne}, A.~G. and {Madsen}, E.~C. and {McLaughlin}, M.~A. and {Patel}, C. and {Scholz}, P. and {Stairs}, I.~H. and {Stappers}, B.~W. and {Zhu}, W.~W.},
        title = "{Eight Millisecond Pulsars Discovered in the Arecibo PALFA Survey}",
      journal = {\apj},
         year = 2019,
        month = dec,
       volume = {886},
       number = {2},
          eid = {148},
        pages = {148},
          doi = {10.3847/1538-4357/ab4f85},
archivePrefix = {arXiv},
       eprint = {1908.09926},
 primaryClass = {astro-ph.HE},
       adsurl = {https://ui.adsabs.harvard.edu/abs/2019ApJ...886..148P}
}

@ARTICLE{Tauris2012,
       author = {{Tauris}, T.~M. and {Langer}, N. and {Kramer}, M.},
        title = "{Formation of millisecond pulsars with CO white dwarf companions - II. Accretion, spin-up, true ages and comparison to MSPs with He white dwarf companions}",
      journal = {\mnras},
         year = 2012,
        month = sep,
       volume = {425},
       number = {3},
        pages = {1601-1627},
          doi = {10.1111/j.1365-2966.2012.21446.x},
archivePrefix = {arXiv},
       eprint = {1206.1862},
 primaryClass = {astro-ph.SR},
       adsurl = {https://ui.adsabs.harvard.edu/abs/2012MNRAS.425.1601T}
}

@ARTICLE{Paczynski1971,
       author = {{Paczy{\'n}ski}, B.},
        title = "{Evolutionary Processes in Close Binary Systems}",
      journal = {\araa},
         year = 1971,
        month = jan,
       volume = {9},
        pages = {183},
          doi = {10.1146/annurev.aa.09.090171.001151},
       adsurl = {https://ui.adsabs.harvard.edu/abs/1971ARA&A...9..183P}
}

@INPROCEEDINGS{Paczynski1976,
       author = {{Paczynski}, B.},
        title = "{Common Envelope Binaries}",
    booktitle = {Structure and Evolution of Close Binary Systems},
         year = 1976,
       editor = {{Eggleton}, Peter and {Mitton}, Simon and {Whelan}, John},
       series = {Proc. IAU Symp.},
       volume = {73},
        month = jan,
        pages = {75},
        publisher= {Reidel, Dordrecht},
       adsurl = {https://ui.adsabs.harvard.edu/abs/1976IAUS...73...75P}
}

@ARTICLE{Ivanova2013,
       author = {{Ivanova}, N. and {Justham}, S. and {Chen}, X. and {De Marco}, O. and {Fryer}, C.~L. and {Gaburov}, E. and {Ge}, H. and {Glebbeek}, E. and {Han}, Z. and {Li}, X. -D. and {Lu}, G. and {Marsh}, T. and {Podsiadlowski}, P. and {Potter}, A. and {Soker}, N. and {Taam}, R. and {Tauris}, T.~M. and {van den Heuvel}, E.~P.~J. and {Webbink}, R.~F.},
        title = "{Common envelope evolution: where we stand and how we can move forward}",
      journal = {\aapr},
         year = 2013,
        month = feb,
       volume = {21},
          eid = {59},
        pages = {59},
          doi = {10.1007/s00159-013-0059-2},
archivePrefix = {arXiv},
       eprint = {1209.4302},
 primaryClass = {astro-ph.HE},
       adsurl = {https://ui.adsabs.harvard.edu/abs/2013A&ARv..21...59I}
}

@ARTICLE{IbenLivio1993,
       author = {{Iben}, Icko, Jr. and {Livio}, Mario},
        title = "{Common Envelopes in Binary Star Evolution}",
      journal = {\pasp},
         year = 1993,
        month = dec,
       volume = {105},
        pages = {1373},
          doi = {10.1086/133321},
       adsurl = {https://ui.adsabs.harvard.edu/abs/1993PASP..105.1373I}
}

@ARTICLE{Rappaport1995,
       author = {{Rappaport}, S. and {Podsiadlowski}, Ph. and {Joss}, P.~C. and {Di Stefano}, R. and {Han}, Z.},
        title = "{The relation between white dwarf mass and orbital period in wide binary radio pulsars}",
      journal = {\mnras},
         year = 1995,
        month = apr,
       volume = {273},
       number = {3},
        pages = {731-741},
          doi = {10.1093/mnras/273.3.731},
       adsurl = {https://ui.adsabs.harvard.edu/abs/1995MNRAS.273..731R}
}

@ARTICLE{TaurisSavonije99,
       author = {{Tauris}, Thomas M. and {Savonije}, Gerrit J.},
        title = "{Formation of millisecond pulsars. I. Evolution of low-mass X-ray binaries with P\_orb> 2 days}",
      journal = {\aap},
         year = 1999,
        month = oct,
       volume = {350},
        pages = {928-944},
archivePrefix = {arXiv},
       eprint = {astro-ph/9909147},
 primaryClass = {astro-ph},
       adsurl = {https://ui.adsabs.harvard.edu/abs/1999A&A...350..928T}
}

@ARTICLE{Tauris2011,
       author = {{Tauris}, T.~M. and {Langer}, N. and {Kramer}, M.},
        title = "{Formation of millisecond pulsars with CO white dwarf companions - I. PSR J1614-2230: evidence for a neutron star born massive}",
      journal = {\mnras},
         year = 2011,
        month = sep,
       volume = {416},
       number = {3},
        pages = {2130-2142},
          doi = {10.1111/j.1365-2966.2011.19189.x},
archivePrefix = {arXiv},
       eprint = {1103.4996},
 primaryClass = {astro-ph.SR},
       adsurl = {https://ui.adsabs.harvard.edu/abs/2011MNRAS.416.2130T}
}

@ARTICLE{vanDenHeuvel994,
       author = {\VAN{Van den Heuvel}{van den Heuvel}{van den Heuvel}, E.~P.~J.},
        title = "{The binary pulsar PSRJ2145-0750: a system originating from a low or intermediate mass X-ray binary with a donor star on the asymptotic giant branch?}",
      journal = {\aap},
         year = 1994,
        month = nov,
       volume = {291},
        pages = {L39-L42},
       adsurl = {https://ui.adsabs.harvard.edu/abs/1994A&A...291L..39V}
}

@INPROCEEDINGS{Tauris2011conf,
       author = {{Tauris}, T.~M.},
        title = "{Five and a Half Roads to Form a Millisecond Pulsar}",
    booktitle = {Evolution of Compact Binaries},
         year = 2011,
       editor = {{Schmidtobreick}, L. and {Schreiber}, M.~R. and {Tappert}, C.},
       series = {Astronomical Society of the Pacific Conference Series},
       volume = {447},
        month = sep,
        pages = {285},
          doi = {10.48550/arXiv.1106.0897},
archivePrefix = {arXiv},
       eprint = {1106.0897},
 primaryClass = {astro-ph.HE},
       adsurl = {https://ui.adsabs.harvard.edu/abs/2011ASPC..447..285T}
}

@ARTICLE{Wang2025,
       author = {{Wang}, P.~F. and {Han}, J.~L. and {Yang}, Z.~L. and {Wang}, T. and {Wang}, C. and {Su}, W.~Q. and {Xu}, J. and {Zhou}, D.~J. and {Yan}, Yi and {Jing}, W.~C. and {Cai}, N.~N. and {Yuan}, J.~P. and {Xu}, R.~X. and {Wang}, H.~G. and {You}, X.~P.},
        title = "{The FAST Galactic Plane Pulsar Snapshot Survey. VIII. 116 Binary Pulsars}",
      journal = {Research in Astronomy and Astrophysics},
         year = 2025,
        month = jan,
       volume = {25},
       number = {1},
          eid = {014003},
        pages = {014003},
          doi = {10.1088/1674-4527/ada3b8},
archivePrefix = {arXiv},
       eprint = {2412.03062},
 primaryClass = {astro-ph.HE},
       adsurl = {https://ui.adsabs.harvard.edu/abs/2025RAA....25a4003W}
}

@ARTICLE{RefsdalWeigert1970,
       author = {{Refsdal}, S. and {Weigert}, A.},
        title = "{Shell Source Burning Stars with Highly Condensed Cores}",
      journal = {\aap},
         year = 1970,
        month = jul,
       volume = {6},
        pages = {426},
       adsurl = {https://ui.adsabs.harvard.edu/abs/1970A&A.....6..426R}
}

@BOOK{Kippenhahn2012,
       author = {{Kippenhahn}, Rudolf and {Weigert}, Alfred and {Weiss}, Achim},
        title = "{Stellar Structure and Evolution}",
         year = 2012,
          doi = {10.1007/978-3-642-30304-3},
          publisher = {Springer},
       adsurl = {https://ui.adsabs.harvard.edu/abs/2012sse..book.....K}
}

@ARTICLE{SC1942,
       author = {{Sch{\"o}nberg}, M. and {Chandrasekhar}, S.},
        title = "{On the Evolution of the Main-Sequence Stars.}",
      journal = {\apj},
         year = 1942,
        month = sep,
       volume = {96},
        pages = {161},
          doi = {10.1086/144444},
       adsurl = {https://ui.adsabs.harvard.edu/abs/1942ApJ....96..161S}
}

@ARTICLE{MillerBertolami2022,
       author = {{Miller Bertolami}, Marcelo M.},
        title = "{A Red Giants' Toy Story}",
      journal = {\apj},
         year = 2022,
        month = dec,
       volume = {941},
       number = {2},
          eid = {149},
        pages = {149},
          doi = {10.3847/1538-4357/ac98c1},
archivePrefix = {arXiv},
       eprint = {2210.07005},
 primaryClass = {astro-ph.SR},
       adsurl = {https://ui.adsabs.harvard.edu/abs/2022ApJ...941..149M}
}

@ARTICLE{Paxton2011,
       author = {{Paxton}, Bill and {Bildsten}, Lars and {Dotter}, Aaron and {Herwig}, Falk and {Lesaffre}, Pierre and {Timmes}, Frank},
        title = "{Modules for Experiments in Stellar Astrophysics (MESA)}",
      journal = {\apjs},
         year = 2011,
        month = jan,
       volume = {192},
       number = {1},
          eid = {3},
        pages = {3},
          doi = {10.1088/0067-0049/192/1/3},
archivePrefix = {arXiv},
       eprint = {1009.1622},
 primaryClass = {astro-ph.SR},
       adsurl = {https://ui.adsabs.harvard.edu/abs/2011ApJS..192....3P}
}

@ARTICLE{Paxton2013,
       author = {{Paxton}, Bill and {Cantiello}, Matteo and {Arras}, Phil and {Bildsten}, Lars and {Brown}, Edward F. and {Dotter}, Aaron and {Mankovich}, Christopher and {Montgomery}, M.~H. and {Stello}, Dennis and {Timmes}, F.~X. and {Townsend}, Richard},
        title = "{Modules for Experiments in Stellar Astrophysics (MESA): Planets, Oscillations, Rotation, and Massive Stars}",
      journal = {\apjs},
         year = 2013,
        month = sep,
       volume = {208},
       number = {1},
          eid = {4},
        pages = {4},
          doi = {10.1088/0067-0049/208/1/4},
archivePrefix = {arXiv},
       eprint = {1301.0319},
 primaryClass = {astro-ph.SR},
       adsurl = {https://ui.adsabs.harvard.edu/abs/2013ApJS..208....4P}
}

@ARTICLE{Paxton2015,
       author = {{Paxton}, Bill and {Marchant}, Pablo and {Schwab}, Josiah and {Bauer}, Evan B. and {Bildsten}, Lars and {Cantiello}, Matteo and {Dessart}, Luc and {Farmer}, R. and {Hu}, H. and {Langer}, N. and {Townsend}, R.~H.~D. and {Townsley}, Dean M. and {Timmes}, F.~X.},
        title = "{Modules for Experiments in Stellar Astrophysics (MESA): Binaries, Pulsations, and Explosions}",
      journal = {\apjs},
         year = 2015,
        month = sep,
       volume = {220},
       number = {1},
          eid = {15},
        pages = {15},
          doi = {10.1088/0067-0049/220/1/15},
archivePrefix = {arXiv},
       eprint = {1506.03146},
 primaryClass = {astro-ph.SR},
       adsurl = {https://ui.adsabs.harvard.edu/abs/2015ApJS..220...15P}
}

@ARTICLE{Paxton2018,
       author = {{Paxton}, Bill and {Schwab}, Josiah and {Bauer}, Evan B. and {Bildsten}, Lars and {Blinnikov}, Sergei and {Duffell}, Paul and {Farmer}, R. and {Goldberg}, Jared A. and {Marchant}, Pablo and {Sorokina}, Elena and {Thoul}, Anne and {Townsend}, Richard H.~D. and {Timmes}, F.~X.},
        title = "{Modules for Experiments in Stellar Astrophysics (MESA): Convective Boundaries, Element Diffusion, and Massive Star Explosions}",
      journal = {\apjs},
         year = 2018,
        month = feb,
       volume = {234},
       number = {2},
          eid = {34},
        pages = {34},
          doi = {10.3847/1538-4365/aaa5a8},
archivePrefix = {arXiv},
       eprint = {1710.08424},
 primaryClass = {astro-ph.SR},
       adsurl = {https://ui.adsabs.harvard.edu/abs/2018ApJS..234...34P}
}

@ARTICLE{Paxton2019,
       author = {{Paxton}, Bill and {Smolec}, R. and {Schwab}, Josiah and {Gautschy}, A. and {Bildsten}, Lars and {Cantiello}, Matteo and {Dotter}, Aaron and {Farmer}, R. and {Goldberg}, Jared A. and {Jermyn}, Adam S. and {Kanbur}, S.~M. and {Marchant}, Pablo and {Thoul}, Anne and {Townsend}, Richard H.~D. and {Wolf}, William M. and {Zhang}, Michael and {Timmes}, F.~X.},
        title = "{Modules for Experiments in Stellar Astrophysics (MESA): Pulsating Variable Stars, Rotation, Convective Boundaries, and Energy Conservation}",
      journal = {\apjs},
         year = 2019,
        month = jul,
       volume = {243},
       number = {1},
          eid = {10},
        pages = {10},
          doi = {10.3847/1538-4365/ab2241},
archivePrefix = {arXiv},
       eprint = {1903.01426},
 primaryClass = {astro-ph.SR},
       adsurl = {https://ui.adsabs.harvard.edu/abs/2019ApJS..243...10P}
}

@ARTICLE{Jermyn2023,
       author = {{Jermyn}, Adam S. and {Bauer}, Evan B. and {Schwab}, Josiah and {Farmer}, R. and {Ball}, Warrick H. and {Bellinger}, Earl P. and {Dotter}, Aaron and {Joyce}, Meridith and {Marchant}, Pablo and {Mombarg}, Joey S.~G. and {Wolf}, William M. and {Sunny Wong}, Tin Long and {Cinquegrana}, Giulia C. and {Farrell}, Eoin and {Smolec}, R. and {Thoul}, Anne and {Cantiello}, Matteo and {Herwig}, Falk and {Toloza}, Odette and {Bildsten}, Lars and {Townsend}, Richard H.~D. and {Timmes}, F.~X.},
        title = "{Modules for Experiments in Stellar Astrophysics (MESA): Time-dependent Convection, Energy Conservation, Automatic Differentiation, and Infrastructure}",
      journal = {\apjs},
         year = 2023,
        month = mar,
       volume = {265},
       number = {1},
          eid = {15},
        pages = {15},
          doi = {10.3847/1538-4365/acae8d},
archivePrefix = {arXiv},
       eprint = {2208.03651},
 primaryClass = {astro-ph.SR},
       adsurl = {https://ui.adsabs.harvard.edu/abs/2023ApJS..265...15J}
}

@ARTICLE{Eggleton1983,
       author = {{Eggleton}, P.~P.},
        title = "{Aproximations to the radii of Roche lobes.}",
      journal = {\apj},
         year = 1983,
        month = may,
       volume = {268},
        pages = {368-369},
          doi = {10.1086/160960},
       adsurl = {https://ui.adsabs.harvard.edu/abs/1983ApJ...268..368E}
}

@ARTICLE{RefsdalWeigert1969,
       author = {{Refsdal}, S. and {Weigert}, A.},
        title = "{Evolution with mass exchange in close binary systems of total mass 2.5 M sun.}",
      journal = {\aap},
         year = 1969,
        month = feb,
       volume = {1},
        pages = {167},
       adsurl = {https://ui.adsabs.harvard.edu/abs/1969A&A.....1..167R}
}

@ARTICLE{RefsdalWeigert1971,
       author = {{Refsdal}, S. and {Weigert}, A.},
        title = "{On the Production of White Dwarfs in Binary Systems of Small Mass}",
      journal = {\aap},
         year = 1971,
        month = aug,
       volume = {13},
        pages = {367},
       adsurl = {https://ui.adsabs.harvard.edu/abs/1971A&A....13..367R}
}

@ARTICLE{OzelFreire2016,
       author = {{{\"O}zel}, Feryal and {Freire}, Paulo},
        title = "{Masses, Radii, and the Equation of State of Neutron Stars}",
      journal = {\araa},
         year = 2016,
        month = sep,
       volume = {54},
        pages = {401-440},
          doi = {10.1146/annurev-astro-081915-023322},
archivePrefix = {arXiv},
       eprint = {1603.02698},
 primaryClass = {astro-ph.HE},
       adsurl = {https://ui.adsabs.harvard.edu/abs/2016ARA&A..54..401O}
}

@ARTICLE{PodsiadlowskiRappaport2000,
       author = {{Podsiadlowski}, Ph. and {Rappaport}, S.},
        title = "{Cygnus X-2: The Descendant of an Intermediate-Mass X-Ray Binary}",
      journal = {\apj},
         year = 2000,
        month = feb,
       volume = {529},
       number = {2},
        pages = {946-951},
          doi = {10.1086/308323},
archivePrefix = {arXiv},
       eprint = {astro-ph/9906045},
 primaryClass = {astro-ph},
       adsurl = {https://ui.adsabs.harvard.edu/abs/2000ApJ...529..946P}
}

@ARTICLE{KingRitter1999,
       author = {{King}, A.~R. and {Ritter}, H.},
        title = "{Cygnus X-2, super-Eddington mass transfer, and pulsar binaries}",
      journal = {\mnras},
         year = 1999,
        month = oct,
       volume = {309},
       number = {1},
        pages = {253-260},
          doi = {10.1046/j.1365-8711.1999.02862.x},
archivePrefix = {arXiv},
       eprint = {astro-ph/9812343},
 primaryClass = {astro-ph},
       adsurl = {https://ui.adsabs.harvard.edu/abs/1999MNRAS.309..253K}
}

@ARTICLE{Joss1987,
       author = {{Joss}, P.~C. and {Rappaport}, S. and {Lewis}, W.},
        title = "{The Core Mass--Radius Relation for Giants: A New Test of Stellar Evolution Theory}",
      journal = {\apj},
         year = 1987,
        month = aug,
       volume = {319},
        pages = {180},
          doi = {10.1086/165443},
       adsurl = {https://ui.adsabs.harvard.edu/abs/1987ApJ...319..180J}
}

@ARTICLE{Savonije1987,
       author = {{Savonije}, G.~J.},
        title = "{A determination of the white-dwarf masses in wide binary radio-pulsar systems}",
      journal = {\nat},
         year = 1987,
        month = jan,
       volume = {325},
       number = {6103},
        pages = {416-418},
          doi = {10.1038/325416a0},
       adsurl = {https://ui.adsabs.harvard.edu/abs/1987Natur.325..416S}
}

@ARTICLE{GlanzPerets2021,
       author = {{Glanz}, Hila and {Perets}, Hagai B.},
        title = "{Common envelope evolution of eccentric binaries}",
      journal = {\mnras},
         year = 2021,
        month = oct,
       volume = {507},
       number = {2},
        pages = {2659-2670},
          doi = {10.1093/mnras/stab2291},
archivePrefix = {arXiv},
       eprint = {2105.02227},
 primaryClass = {astro-ph.SR},
       adsurl = {https://ui.adsabs.harvard.edu/abs/2021MNRAS.507.2659G}
}

@ARTICLE{Szolgyen2022,
       author = {{Sz{\"o}lgy{\'e}n}, {\'A}kos and {MacLeod}, Morgan and {Loeb}, Abraham},
        title = "{Eccentricity evolution in gaseous dynamical friction}",
      journal = {\mnras},
         year = 2022,
        month = jul,
       volume = {513},
       number = {4},
        pages = {5465-5473},
          doi = {10.1093/mnras/stac1294},
archivePrefix = {arXiv},
       eprint = {2203.01334},
 primaryClass = {astro-ph.EP},
       adsurl = {https://ui.adsabs.harvard.edu/abs/2022MNRAS.513.5465S}
}

@ARTICLE{Trani2022,
       author = {{Trani}, Alessandro Alberto and {Rieder}, Steven and {Tanikawa}, Ataru and {Iorio}, Giuliano and {Martini}, Riccardo and {Karelin}, Georgii and {Glanz}, Hila and {Portegies Zwart}, Simon},
        title = "{Revisiting the common envelope evolution in binary stars: A new semianalytic model for N -body and population synthesis codes}",
      journal = {\prd},
         year = 2022,
        month = aug,
       volume = {106},
       number = {4},
          eid = {043014},
        pages = {043014},
          doi = {10.1103/PhysRevD.106.043014},
archivePrefix = {arXiv},
       eprint = {2205.13537},
 primaryClass = {astro-ph.SR},
       adsurl = {https://ui.adsabs.harvard.edu/abs/2022PhRvD.106d3014T}
}

@ARTICLE{Rappaport1982,
       author = {{Rappaport}, S. and {Joss}, P.~C. and {Webbink}, R.~F.},
        title = "{The evolution of highly compact binary stellar systems.}",
      journal = {\apj},
         year = 1982,
        month = mar,
       volume = {254},
        pages = {616-640},
          doi = {10.1086/159772},
       adsurl = {https://ui.adsabs.harvard.edu/abs/1982ApJ...254..616R}
}

@ARTICLE{Temmink2023,
       author = {{Temmink}, K.~D. and {Pols}, O.~R. and {Justham}, S. and {Istrate}, A.~G. and {Toonen}, S.},
        title = "{Coping with loss. Stability of mass transfer from post-main-sequence donor stars}",
      journal = {\aap},
         year = 2023,
        month = jan,
       volume = {669},
          eid = {A45},
        pages = {A45},
          doi = {10.1051/0004-6361/202244137},
archivePrefix = {arXiv},
       eprint = {2209.12707},
 primaryClass = {astro-ph.SR},
       adsurl = {https://ui.adsabs.harvard.edu/abs/2023A&A...669A..45T}
}

@ARTICLE{Manchester2005,
       author = {{Manchester}, R.~N. and {Hobbs}, G.~B. and {Teoh}, A. and {Hobbs}, M.},
        title = "{The Australia Telescope National Facility Pulsar Catalogue}",
      journal = {\aj},
         year = "2005",
        month = "Apr",
       volume = {129},
       number = {4},
        pages = {1993-2006},
          doi = {10.1086/428488},
archivePrefix = {arXiv},
       eprint = {astro-ph/0412641},
 primaryClass = {astro-ph},
       adsurl = {https://ui.adsabs.harvard.edu/abs/2005AJ....129.1993M}
}

@ARTICLE{RasioHeggie95,
       author = {{Rasio}, Frederic A. and {Heggie}, Douglas C.},
        title = "{The Orbital Eccentricities of Binary Millisecond Pulsars in Globular Clusters}",
      journal = {\apjl},
         year = 1995,
        month = jun,
       volume = {445},
        pages = {L133},
          doi = {10.1086/187907},
archivePrefix = {arXiv},
       eprint = {astro-ph/9502105},
 primaryClass = {astro-ph},
       adsurl = {https://ui.adsabs.harvard.edu/abs/1995ApJ...445L.133R}
}

@ARTICLE{Zahn1977,
       author = {{Zahn}, J.-P.},
        title = "{Tidal friction in close binary systems.}",
      journal = {\aap},
         year = 1977,
        month = may,
       volume = {57},
        pages = {383-394},
       adsurl = {https://ui.adsabs.harvard.edu/abs/1977A&A....57..383Z}
}

@ARTICLE{Nie2026,
       author = {{Nie}, Yu-Dong and {Shao}, Yong and {He}, Jian-Guo and {Wei}, Ze-Lin and {Gao}, Shi-Jie and {Xu}, Xiao-Jie and {Li}, Xiang-Dong},
        title = "{Formation of Recycled Pulsars in Common Envelope Binaries}",
      journal = {\apj},
         year = 2026,
        month = feb,
       volume = {997},
       number = {2},
          eid = {265},
        pages = {265},
          doi = {10.3847/1538-4357/ae3569},
archivePrefix = {arXiv},
       eprint = {2601.04355},
 primaryClass = {astro-ph.HE},
       adsurl = {https://ui.adsabs.harvard.edu/abs/2026ApJ...997..265N}
}

@ARTICLE{Brown2020,
       author = {{Brown}, Warren R. and {Kilic}, Mukremin and {Kosakowski}, Alekzander and {Andrews}, Jeff J. and {Heinke}, Craig O. and {Ag{\"u}eros}, Marcel A. and {Camilo}, Fernando and {Gianninas}, A. and {Hermes}, J.~J. and {Kenyon}, Scott J.},
        title = "{The ELM Survey. VIII. Ninety-eight Double White Dwarf Binaries}",
      journal = {\apj},
         year = 2020,
        month = jan,
       volume = {889},
       number = {1},
          eid = {49},
        pages = {49},
          doi = {10.3847/1538-4357/ab63cd},
archivePrefix = {arXiv},
       eprint = {2002.00064},
 primaryClass = {astro-ph.SR},
       adsurl = {https://ui.adsabs.harvard.edu/abs/2020ApJ...889...49B}
}

@ARTICLE{SunArras2018,
       author = {{Sun}, M. and {Arras}, P.},
        title = "{Formation of Extremely Low-mass White Dwarf Binaries}",
      journal = {\apj},
         year = 2018,
        month = may,
       volume = {858},
       number = {1},
          eid = {14},
        pages = {14},
          doi = {10.3847/1538-4357/aab9a4},
archivePrefix = {arXiv},
       eprint = {1703.01648},
 primaryClass = {astro-ph.SR},
       adsurl = {https://ui.adsabs.harvard.edu/abs/2018ApJ...858...14S}
}

@ARTICLE{Li2019,
       author = {{Li}, Zhenwei and {Chen}, Xuefei and {Chen}, Hai-Liang and {Han}, Zhanwen},
        title = "{Formation of Extremely Low-mass White Dwarfs in Double Degenerates}",
      journal = {\apj},
         year = 2019,
        month = feb,
       volume = {871},
       number = {2},
          eid = {148},
        pages = {148},
          doi = {10.3847/1538-4357/aaf9a1},
archivePrefix = {arXiv},
       eprint = {1812.07226},
 primaryClass = {astro-ph.SR},
       adsurl = {https://ui.adsabs.harvard.edu/abs/2019ApJ...871..148L}
}







\bsp	
\label{lastpage}
\end{document}